\documentclass{jfm}

\usepackage{microtype}
\usepackage{subcaption} 
\usepackage{graphicx}
\usepackage{newtxtext}
\usepackage{newtxmath}
\usepackage{empheq}

\usepackage{natbib}
\usepackage{hyperref}
\usepackage{enumerate}
\usepackage{enumitem}
\usepackage[usenames, dvipsnames]{xcolor}
\hypersetup{
    colorlinks = true,
linkcolor=MidnightBlue,
    urlcolor   = blue,
    citecolor  = MidnightBlue,
}

\usepackage{mdframed}

\usepackage{multirow}
\usepackage{multicol}

\usepackage[normalem]{ulem}

\usepackage{cleveref}
\crefname{figure}{figure}{figures}
\crefname{appendix}{Appendix}{Appendices}
\crefname{section}{\S}{\S}

\usepackage{siunitx}
\usepackage{booktabs}

\usepackage{pdflscape}
\usepackage{afterpage}
\usepackage{flafter}
\usepackage{rotating}
\usepackage{fancyhdr}
\fancypagestyle{lscape}{%
\fancyhf{} 
\fancyfoot[LE]{}
\fancyfoot[LO] {}
 
}

\newcommand*{\de}{\operatorname{d\!}{}} 
\newcommand{\dd}[2]{\frac{\de#1}{\de#2}}
\newcommand{\pd}[2]{\frac{\partial#1}{\partial#2}}

\newcommand*{\Omegabw}{\Omega_\text{b. water}}
\newcommand*{\Omegai}{\Omega_\text{ice}}
\newcommand*{\Omegatw}{\Omega_\text{t. water}}

\newcommand*{\zwit}{z_\text{w}}
\newcommand*{\ziwt}{z_\text{wi}}
\newcommand*{\ztop}{z_\text{top}}

\newcommand*{\m}{\dot{m}}
\newcommand*{\Trec}{T_\text{rec}}
\newcommand*{\Tsubs}{T_\text{subs}} 

\newcommand*{\kw}{k_\text{w}}
\newcommand*{\ki}{k_\text{i}}
\newcommand*{\cw}{c_\text{w}}
\newcommand*{\ci}{c_\text{i}}
\newcommand*{\rhow}{\rho_\text{w}}
\newcommand*{\rhoi}{\rho_\text{i}}

\newcommand*{\Pe}{\textrm{Pe}}
\newcommand*{\St}{\textrm{St}}
\newcommand*{\Steff}{\textrm{St}_\text{eff}}
\newcommand*{\MR}{M_{r}}
\newcommand*{\D}{D}

\newcommand*{\Lh}{L} 
\newcommand*{\Hc}{H} 
\newcommand*{\K}{K} 
\newcommand*{\Ro}{R} 

\newcommand*{\Ti}{T_\text{ice}}
\newcommand*{\Tw}{T_\text{water}}
\newcommand*{\Tsurf}{T_\text{surf}}
\newcommand*{\Tfrz}{T_\text{frz}}
\newcommand*{\hi}{h_\text{ice}} 
\newcommand*{\hw}{h_\text{water}}
\newcommand*{\hsurf}{h_\text{surf}}
\newcommand*{\htotal}{h_\text{total}}
\newcommand*{\hm}{h_\text{mush}}

\newcommand{\nd}[1]{#1 } 
\newcommand{\ndd}[1]{#1 '} 

\newcommand*{\mimp}{\dot{m}_\text{imp}}
\newcommand*{\mev}{\dot{m}_\text{ev}}
\newcommand*{\Eimp}{E_\text{imp}}

\newcommand*{\Estarp}{\nd{E}^{*}}

\newcommand*{\mf}{\dot{m}_\text{f}}
\newcommand*{\evI}{\alpha_1}
\newcommand*{\evII}{\alpha_2}

\graphicspath{ {images/} }

\shorttitle{An enthalpy model for ice-crystal icing}
\shortauthor{Peters, Shelton, Tang, and Trinh}

\title{An enthalpy-based model for the physics of ice crystal icing}

\author{Timothy Peters\aff{1}, 
 Josh Shelton\aff{1}, Hui Tang\aff{2} \corresp{\email{h.tang2@bath.ac.uk}}
 \and Philippe H. Trinh\aff{1} \corresp{\email{p.trinh@bath.ac.uk}}
 }

\affiliation{\aff{1}Department of Mathematical Sciences, University of Bath, Bath BA2 7AY, UK
\aff{2}Department of Mechanical Engineering, University of Bath, Bath BA2 7AY, UK}

\date{\today~[Draft]}
\AtBeginDocument{}
\begin{document}

\maketitle
\begin{abstract}
Ice crystal icing (ICI) in aircraft engines is a major threat to flight safety. Due to the complex thermodynamic and phase-change conditions involved in ICI, rigorous modelling of the accretion process remains limited. The present study proposes a novel modelling approach based on the physically-observed mixed-phase nature of the accretion layers. The mathematical model, which is derived from the enthalpy change after accretion (the enthalpy model), is compared to an existing pure-phase layer model (the three-layer model). Scaling laws and asymptotic solutions are developed for both models. The onset of ice accretion, the icing layer thickness, and solid ice fraction within the layer are determined by a set of non-dimensional parameters including the P\'{e}clet number, the Stefan number, the Biot number, the Melt Ratio, and the evaporative rate. Thresholds for freezing and non-freezing conditions are developed. The asymptotic solutions presents good agreement with numerical solutions at low P\'{e}clet numbers. Both the asymptotic and numerical solutions show that, when compared to the three-layer model, the enthalpy model presents a thicker icing layer and a thicker water layer above the substrate due to mixed-phased features and modified Stefan conditions. Modelling in terms of the enthalpy poses significant advantages in the development of numerical methods to complex three-dimensional geometrical and flow configurations. These results improve understanding of the accretion process and provide a novel, rigorous mathematical framework for accurate modelling of ICI.
\end{abstract}

\section{\label{sec:intro}Introduction}

\noindent Previously, research on aircraft icing was focused on the freezing of supercooled water droplets on the exterior of aircraft or on a sub-freezing engine surface. Since the mid 1990s, more than 200 engine power loss events were documented at altitudes of more than 7000m where there is hardly any supercooled water content. In 2006, Mason \emph{et al.} attributed these events to the ingestion of ice crystals generated by thunderstorms or convective storms \citep{Mason2006}.  Under subfreezing conditions, ice crystals normally bounce off cold surfaces, and are relatively unproblematic; however, within an aeroengine compressor, the air temperature increases to above freezing. After ingestion into the compressor, ice crystals can melt, stick, and accrete on the interior warm metal surfaces, leading eventually to flow blockage and subsequent engine power loss, stall, and surge. Shedding of accumulated ice further damages engine components, promoting engine failure~\citep{Mason2006}. Therefore, it has been recognised within the aeroengine industry that a deep understanding of the mechanisms and consequences of ice crystal icing (ICI), including the accretion processes, is vital to ensure flight safety \citep{yamazakiReviewCurrentStatus2021}. Such understanding is crucial for implementation of safety protocols, and provision of pilot guidance and aircraft certification requirements, and in underpinning the development of new technologies that can mitigate ice accretion and subsequent engine damage.

Experimental work performed in e.g. the NASA Glenn Research Center \citep{Currie2012,Currie2014}, RATFac Canada \citep{Bucknell2019Experimental}, and Icing Wind Tunnel Braunschweig \citep{Baumert2018Experimental} has improved the understanding of the physical mechanisms of the ice crystal icing. It was shown that the accretion of mixed-phase water content was dependent on the ice fraction of the impinging water content, the ambient temperature and humidity conditions.

Numerous models have been adapted in order to account for ice crystal accretion, including \citet{Kintea2014,Kintea2016,Tsao2016,bartkus2018evaluation,Bartkus2019,villedieu2014glaciated,trontin2018comprehensive, bucknell2019three,ayan2018flight}; these are often an extension of the Model proposed by \citet{messinger1953equilibrium}.
More recent models that are based on the extended Messinger models (EMM) are by ~\citet{ayan2018flight} which allowed for mixed phase and glaciated conditions but only for sub-freezing temperatures, as well as \citet{bucknell2019three} who adapted the EMM for above freezing temperatures. A summary of previous works on ice crystal icing can be found in \cref{tab:ref_guide}. 

\citet{bucknell2019three} developed a three layer thermodynamic model for ice accretion, which accounts for warm surfaces and mixed phase conditions. In the case of a warm substrate, the model is split into two regimes: (i) running wet conditions where there exists only a water film and all impinging ice is melted; (ii) if the water layer surface temperature reaches freezing, an ice layer forms over the water film and is in turn below a small surface water film. However, the model is based on the assumption that all the impinging ice in the second regime accumulates into a pure solid ice layer, which is not a representation of the ice accretion physics observed by \citet{Malik2024}. \citet{Malik2024} shows that the accretion layer is a combination of liquid and solid state. Therefore, this paper is to present a novel 1-D model of ice crystal icing with a warm substrate and a mixed-phase icing layer.

\begin{figure}
    \centering
    \includegraphics[scale=0.2]{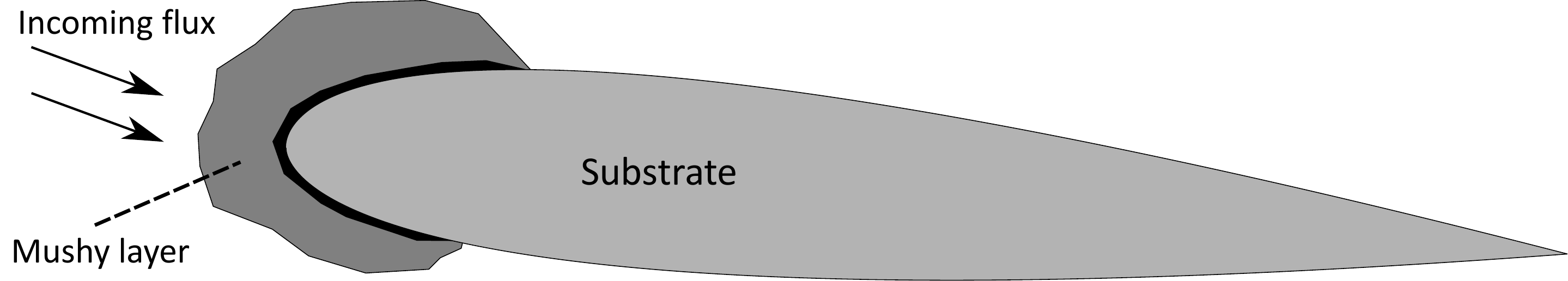}
    \caption{Schematic of the accretion composition for incoming mass flux on a warm substrate as described by the enthalpy model in \S{\ref{sec:dimensionalEnth}}. This consists of a thin water layer on the substrate, with a mixed-phase water/ice (mush) composition on top.
    }
    \label{fig:enter-label}
\end{figure}

At the melting point, the liquid water and ice have the same temperature but different enthalpy. This enthalpy difference is referred to as the latent heat of fusion. The enthalpy level of the mixture can be used to quantify the ice fraction. In addition, the interface between the pure water layer and the mixed-phase layer can be determined from the enthalpy distribution, which, as discussed by \citet{CrankBook1984}, simplifies the simulation of moving boundary problems. Further advances in numerical schemes for enthalpy such as the flag scheme developed by \cite{bridge2007mixture} ensure accuracy with reduced computation cost. Therefore, we study the enthalpy formulation of the problem. In addition, the formulation of the model used by \citet{bucknell2019three} is also studied for comparison. The main differences between the two models are: 

\begin{enumerate}[label=(\roman*),leftmargin=*, align = left, labelsep=\parindent, topsep=3pt, itemsep=2pt,itemindent=0pt ]
\item The three-layer model of \cite{bucknell2019three}, outlined in \S\ref{sec:fixedfront} serves as an extension to the extended Messinger model \citep{myers2001extension} for substrate temperatures above freezing. 
The model initially contains one water layer. When the water surface reaches freezing temperature, it transitions to a three-layer configuration, consisting of water, ice, and water layers. The partially melted ice particles leads to surface accretion. In the presence of just one water layer, it is assumed that this ice immediately melts. When all three layers are present, the surface accretion contributes to both the ice layer and the top water layer.

\item The new enthalpy model, developed in \S\ref{sec:dimensionalEnth}, ultimately leads to a two-layer configuration consisting of a water phase and a mixed water/ice phase (mush). This is a PDE with a single free boundary at the surface. Since the enthalpy is defined as the sum of sensible and latent heats, it is capable of capturing multiple layers, and phase changes, without the specification of interfacial equations. After freezing temperature is reached, the model effectively contains a lower water layer, and an upper mushy layer with mixed-phase properties of ice and water.
\end{enumerate}

\noindent The governing equations are non-dimensionalised, showing that the icing problem is controlled by a substantial group of non-dimensional parameters, including the P\'eclet number $\Pe$, Stefan number St, melt ratio Mr, Biot number Bi, $\m_\text{ev}$, ratio of latent heats L, and the kinetic ratio D. In addition, we shall develop asymptotic approximations in the limit of $\Pe \to 0$; these provide simple expressions for water and ice growth rates for both the three-layer and enthalpy models. The leading order approximation is equivalent to the solution if the heat transfer within the accretion layers is assumed to be quasi-steady (the transient term in heat equations are ignored). The asymptotic approximations are compared to numerical simulations for both models. The difference between the accretion characteristics of both models are compared and discussed, parametric studies are conducted to show the effects of non-dimensional parameters. Our analysis, including the asymptotic approximations, allow clear identification of the role of different parameters in the dynamics of the ice crystal icing problem. Prior works of the general ice-crystal icing problem have not performed asymptotic analysis to this level of detail; as the complexity of the problem increases (to include, \emph{e.g.} three-dimensional dynamics) the asymptotic growth laws developed in this work is useful for benchmarking and validation purposes.

\subsection{Outline of the paper}
The three-layer formulation of water-ice-water is derived in \S\ref{sec:fixedfront}.
We develop our enthalpy model in \S\ref{sec:dimensionalEnth}; this captures mixed-phase regions of water and ice.
The key dimensional and nondimensional parameters are given in \S\ref{sec:parameters}, in which the values used in our models are provided. Asymptotic solutions for small P\'eclet number are developed in \S\ref{sec:asympt}, for both phases of the three-layer model, and for our enthalpy model. Comparisons with numerical solutions are presented in \S\ref{sec:num}. Section~\ref{sec:conclusions} summarises the key conclusions, and further discussion occurs in \S\ref{sec:discussion}.

\begin{table}
    \centering\footnotesize
    \begin{tabular}{p{0.9cm}p{3cm}p{3cm}p{4.5cm}p{1.6cm}}
        \textsc{Type} & \textsc{Reference} & \textsc{Model} & \textsc{Comments}  & \textsc{Substr.} \\[1em] 
        \multirow{10}*{MM} & \cite{villedieu2014glaciated} &  2D with thin film for water runback  & Adaptation of previous icing codes to allow for ice crystals & FT\\
        & \cite{tsao2014possible}& 1D at stagnation point  & Generate early estimates for ice accretion & FT\\ 
        & \cite{Bartkus2019} & 1D at stagnation point  & Introduced thermal coupling between accretion and substrate. Examine losses between prediction and experiment & C\\
        & \cite{nilamdeen2019numerical} & 3D with thin film for water runback  & Examines the effect of vapour and the evaporative mass flux & U \\\hline
        \multirow{10}*{EMM} & \cite{ayan2018flight} & 2D with no water runback & Adapting EMM to allow for ice crystals & FT \\
        & \cite{bucknell2019three}  & 2D with panel method for water runback & Adapting EMM to allow for ice crystals and surfaces above freezing & P\\
        & \cite{connolly2021ice} & 2D with bulk solution of thin film for runback  & Improving on the model by \cite{bucknell2019three}. Adaption of water runback and treatment of substratum & C \\  
        & \cite{gallia2023comprehensive} & 2D with thin film for water runback& Adapting \cite{myers1999ice} to consider the effect of a heated surface and thus shedding effects  & U \\ \hline
        \multirow{15}*{Porous} & & & \\
        & \cite{Kintea2014}  & 3D finite volume with no runback & Examination of substratum cooling and shedding thresholds  & P\\
        & \cite{trontin2018comprehensive}& 2D MM with porous accretion. Empirical relations for water runback. & Updated version of \cite{villedieu2014glaciated} allowing porosity and inclusion of other effects  & FT\\
        & \cite{roychowdhury2023ice} & 2D thin film model of water/ice mixture& Model based off of \cite{malik2022experimental}  & P \\ 
        & \cite{zhang2023three} & 2D ice/mixture with surface thin film for runback & Allows a total void space in the porous medium and employs a wicking rate between the water and mixture. & P\\
        & \cite{malik2023experimental}  & 2D thin film model of water/ice mixture & Considers an ice/water mixture and examines two-way coupling between accretion and substratum & C \\ \hline
        \multirow{5}*{Other} & & & & \\
        & \cite{currie2020physics} & 2D with bulk solution of thin film for runback & Examines how and where ice melts after being trapped in water. Considers the cooling down of the 2D substratum & C \\ \hline
        \end{tabular}
        \begin{tabular}{p{2cm}p{11.5cm}}
        \multirow{2}*{Experiments} & \cite{currie2011fundamental,Currie2012,currie2013altitude,Currie2014,struk2015ice,struk2017initial, struk2018ice, struk2019ice,Baumert2018Experimental, bucknell2018experimental, Bucknell2019Experimental, malik2022experimental}  
    \end{tabular}
    \caption{A selection of different formulations used for modelling ice crystal icing in engines. We introduce the following notation to refer to the substratum (\textsc{substr.}): FT (freezing temperature), C (coupling between the accretion and substratum), P (prescribed heat flux or temperature) and U (unspecified).} 
    \label{tab:ref_guide}
\end{table}

\section{Mathematical formulation of a three-layer model} \label{sec:fixedfront}

Previously, ice crystal icing models which implemented the supercooled droplet model of \cite{messinger1953equilibrium} or the extended Messinger model \citep{myers2001extension} have assumed a substrate temperature that is below or equal to freezing temperature \citep{ayan2018flight}. In such models, ice is the first layer present on the solid surface. This incorrectly models the formation of ice within engines, in which the first layer must be water due to the presence of a warm substrate \citep{bucknell2019icicle}. The three-layer water-ice-water model, developed by \cite{bucknell2019three}, was designed as a further extension to the extended Messinger model, primarily to allow for substrate temperatures above freezing. In this section, we review this latter model, which we solve using a fixed front method in which the interface between the water and ice layers is tracked and must be determined as part of the solution. 
The extended Messinger model is formulated in \S\ref{sec:EMM}, and the modifications required to include a warm substrate are detailed in the water layer formulation of \S\ref{sec:water}, and the water-ice-water layer formulation of \S\ref{sec:three}. We non-dimensionalise the governing system in \S{\ref{sec:ff_nondim}}, and discuss its connection with the enthalpy model of ice crystal icing, developed later in \S\ref{sec:dimensionalEnth}.

\subsection{A review of the Extended Messinger Model}\label{sec:EMM}

\begin{figure}
    \centering
    \includegraphics[scale=1]{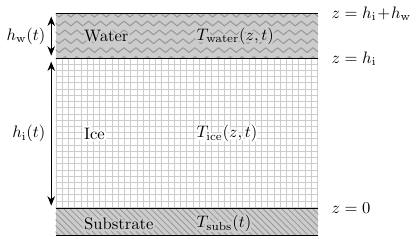}
    \caption{Form of icing captured in the original Messinger model, as depicted in \cite{myers2001extension}.}
    \label{fig:myers_scheme}
\end{figure}

We now introduce the dimensional formulation of the Extended Messinger Model (EMM), for which a typical solution is shown in figure~\ref{fig:myers_scheme}. Solutions of this formulation will depend on the coordinates $z$ and $t$, where $z$ is the spatial coordinate orthogonal to the lower substrate, and $t$ is time. We denote the thickness of the lower ice layer by $\hi(t)$, and the thickness of the upper water layer by $\hw(t)$, both of which will be solved for as part of the solution. The total height of the domain is then given by $z = \hi(t)+ \hw(t)$. The ice temperature, $\Ti(z,t)$, will then be solved for across $0 \leq z \leq \hi(t)$, and the water temperature, $\Tw(z,t)$, will be solved for across $\hi(t) \leq z \leq \hi(t) + \hw(t)$.

The governing equations consist of mass conservation for the total growth of the system:
\begin{subequations}
\begin{equation}\label{eq:EMMcontinuity}
        \rho_\text{w}\dd{\hw}{t} + \rho_\text{i}\dd{\hi}{t} = \m_{\text{imp}}-\dot{m}_{\text{ev,sub}},
    \end{equation}
where $\m_{\text{imp}}$ is the impinging mass flux which can be broken into water and ice contributions; this contribution crucially depends on the \emph{melt ratio}, which is the ratio of liquid water content to total (ice + water) water content in the incoming ice particles \citep{currie2013altitude}. Note that we assume the temperature of the incoming liquid water and ice content stays at the freezing temperature ($T_\text{frz}=0^\circ C$). We also have mass transfer via the evaporative/sublimative flux, $\m_{\text{ev,sub}}$; dependent on the surrounding conditions, this evaporative/sublimative flux can then model the gain or loss of mass [cf. Appendix~\ref{sec:evap}]. Above, $\rho_w$ and $\rho_i$ correspond to the densities of water and ice, respectively. Typical values of these parameters are listed later in table~\ref{tab:physicalconditions}.

The temperatures in the ice and water layers are given by one-dimensional heat equations,
    \begin{align}
\rhoi c_{\text{i}} \frac{\partial \Ti}{\partial t} &=k_{\text{i}}\frac{\partial^2 \Ti}{\partial z^2} && \text{for} \quad 0 \leq z \leq \hi(t), \label{eq:EMMTice} \\
\rhow c_{\text{w}} \frac{\partial \Tw}{\partial t} &=k_{\text{w}}\frac{\partial^2 \Tw}{\partial z^2} && \text{for} \quad \hi(t) \leq z \leq \hi(t)+\hw(t), \label{eq:EMMTwater}
\end{align}
where $\cw, \kw,\ci,\ki$ denote the heat capacity and thermal conductivity of water and ice, and typical values can be found in table~\ref{tab:physicalconditions}. 

In the original model by \cite{messinger1953equilibrium} [and also discussed in the extended model by \cite{myers2001extension}], only supercooled water droplets are assumed, and the substrate, at $z = 0$, is considered to be held at a temperature below the point of freezing. As a result, all incoming water is assumed to freeze, leading to the instantaneous formation of an ice layer with $\Ti < 0$---this is the situation of {\itshape rime ice}. As it concerns \eqref{eq:EMMcontinuity}, only the growth of the ice layer needs to be considered ($\hw = 0$), and we only need to consider the temperature profile in the ice given by \eqref{eq:EMMTice}. 

As noted by \cite{myers2001extension}, in certain situations and under suitable flux conditions, the surface temperature of the ice layer, i.e. $\Ti(\hi(t), t)$, can reach the freezing temperature, thus allowing formation of a water layer on top; this leads to the setup known as {\itshape glaze ice}. In this case, we also need to consider the temperature of the water given by \eqref{eq:EMMTwater}, as well as an additional energy balance on the ice-water interface given by the {\itshape Stefan condition}:
\begin{equation}\label{eq:EMMStefan} 
    \rho_{\text{i}}L_{\text{f}} \frac{\mathrm{d}\hi}{\mathrm{d}t}= -k_{\text{w}} \frac{\partial \Tw}{\partial z} + k_{\text{i}} \frac{\partial \Ti}{\partial z} \qquad \text{at} \quad z = \hi(t), 
\end{equation}
\end{subequations}
where we have introduced the latent heat of fusion given by $L_{\text{f}}$.
Note that nondimensionaling \eqref{eq:EMMTwater} leads to the P\'{e}clet number, 
\begin{equation}\label{eq:Pe}
\Pe = \frac{\rhow \cw [H]^2}{\kw[t]},
\end{equation}
where $[H]$ and $[t]$ correspond to typical length and temporal scales. Then $\Pe$ governs the balance between advective and diffusive effects, and is often assumed to be small. When $\Pe \ll 1$, the problem becomes quasi-steady as the time derivative components in each heat equation become subdominant. Indeed this assumption was used by a number of authors including \cite{bucknell2019three, gallia2023comprehensive}

Finally, there are a number of assumptions that underpin the above model. These include: (i) lateral conduction is neglected; (ii) there is perfect thermal contact between the accretion and the substrate; (iii) the ice-water interfaces are at the freezing temperature; (iv) conduction within the substrate is not considered and the substrate temperature is prescribed.

The model developed by \cite{bucknell2019three}, denoted by the three-layer model, builds on the extended Messinger model discussed above, but considers two stages for the situation of a warm substrate and partially melted impinging water content. The original dimensional form of the model was presented in \cite{bucknell2019three}. Our goal in the next two subsections is to re-interpret the model rigorously and to derive a non-dimensional form.

\subsection{Stage 1 of Bucknell's model for ice-crystal icing (water-only)}\label{sec:water}

\begin{figure}
    \centering
    \includegraphics[scale=1]{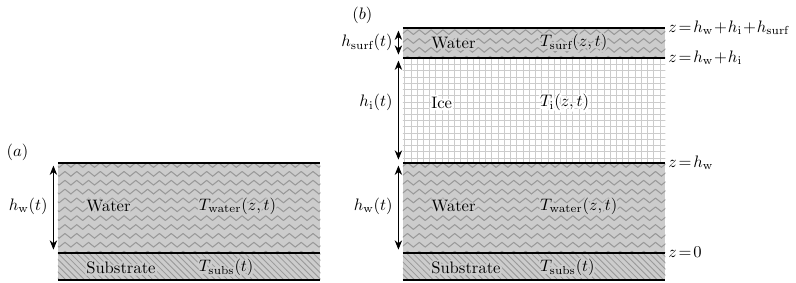}
    \caption{Schematic of the two stages in the three-layer model of ice crystal icing as developed in \cite{bucknell2019three}. Initially, (a) in Stage 1, there is only a water film; then in (b) Stage 2, a water-ice-water layer is used. \label{fig:scheme} }
\end{figure}

In \emph{Stage 1}, all the ice in the impinging water content melts, and leads to the formation of a water layer on the solid substrate. The formation of the initial water layer serves to trap more incoming particles, and thus leads to additional build-up of water. If the energy from melting is not balanced from other contributions, such as via convective and kinetic transport, this leads to a reduction in the water surface temperature. Then, if the temperature at the water surface drops to the freezing temperature, this then brings the model to \emph{Stage 2}, which permits the modelling of both an ice layer and a surface-water layer. These two stages are shown in \cref{fig:scheme}.

\begin{figure}
    \centering
    \includegraphics{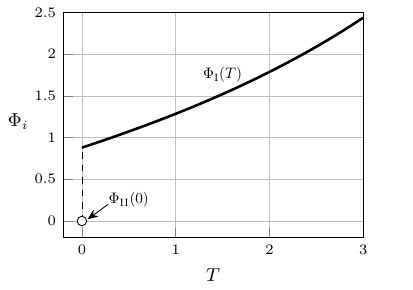}
    \caption{Plot of $T$ vs $\Phi_i(T)$ for $i=\text{I}$ (\S{\ref{sec:water}}), $\text{II}$ (\S{\ref{sec:three}}).}
    \label{fig:PhiTvsT}
\end{figure}

We begin by considering the water-only state, as shown in \cref{fig:scheme}(a). In this first stage, we are interested in the height and temperature of the water. Note that in contrast with the extended Messinger model, the water layer, $\hw(t)$, is now resting on the substrate. We seek to solve:
\begin{subequations}\label{eq:DimBS1}
\begin{align}
\rho c_{\text{w}} \frac{\partial \Tw}{\partial t} &=k_{\text{w}}\frac{\partial^2 \Tw}{\partial z^2} && \text{for} \quad 0 \leq z \leq \hw(t), \label{eq:DimBS1tempwater} \\
\rho \frac{\mathrm{d}\hw}{\mathrm{d}t}&= \dot{m}_{\text{imp}} - \dot{m}_{\text{ev}}(\Tw) && \text{at} \quad z=\hw(t). \label{eq:DimBS1surfaceevolve} 
\end{align}
In practice, \cite{bucknell2019three} use a quasi-static assumption, so the temporal terms in \eqref{eq:DimBS1tempwater} are ignored. The growth of the water layer is given by the continuity equation \eqref{eq:DimBS1surfaceevolve}. 
Our system starts from a clean substrate, and thus we have the following initial condition for the water layer:
\begin{equation*}
    \hw(0) = 0.
\end{equation*}
We then impose boundary conditions for the temperature,
\begin{align} \label{eq:DimBS1substratetemp}
\Tw&=\Tsubs && \text{at} \quad z=0, \\ \label{eq:DimBS1surfaceflux}
-k_{\text{w}}\frac{\partial \Tw}{\partial z}  &=  \Phi_\text{I}(\Tw) && \text{at} \quad z=\hw(t),
\end{align}
where the water adopts the positive substrate temperature, $T_0>0$, on the substrate, and there is a heat flux $\Phi_\text{I}$ on the exposed water surface. The function $\Phi_\text{I}$ can be broken down into components that are dependent on the surface temperature $\Tw(\hw(t),t)$, such as evaporation, convection, and sensible heat fluxes; it should also be considered as functions of components that are independent of $\Tsurf$, such as the melting/freezing and kinetic energies. For the sensible heat flux, we assume that impinging water and ice are at the freezing temperature, $\Tfrz$. Thus, following \cite{bucknell2019three}, we have the following makeup of the flux term:
\end{subequations}
\begin{multline} \label{eq:Phi}
\Phi_\text{I}(\Tw) = \Bigl[\underbrace{\vphantom{\frac{1}{2}}h_{\text{tc}}(\Tw-T_{\text{rec}})}_{\text{convection}}\Bigr] 
+ \Bigl[\underbrace{\vphantom{\frac{1}{2}}L_{\text{v}}\dot{m}_{\text{ev}}(\Tw)}_{\text{evaporation}}\Bigr]
+ \Bigl[\underbrace{\vphantom{\frac{1}{2}}L_{\text{f}}\dot{m}_{\text{imp,i}}}_{\text{melt/freeze}}\Bigr] \\
+ \Bigl[\underbrace{\vphantom{\frac{1}{2}}c_{\text{w}}\dot{m}_{\text{imp}}(\Tw-T_\text{frz})}_{\text{sensible}}\Bigr]
- \Bigl[\underbrace{\frac{1}{2}\dot{m}_{\text{imp}}\bar{U}^2}_{\text{kinetic}}\Bigr].
\end{multline}
We shall later provide typical parameter ranges of the contributions above in \S\ref{sec:ff_nondim}. While complicated, the functional form of $\Phi_{\text{I}}(T)$ in \eqref{eq:Phi} is linear in $T$ for all terms with the exception of the evaporation, $\mev(T)$, which is a nonlinear function. Later, it will be further non-dimensionalised. A typical non-dimensional shape for $\Phi_\text{I}$ is shown in  \cref{fig:PhiTvsT}. Note that at $T=0$, we observe a jump to $\Phi_\text{II}$ as the system enters stage two, described in \S{\ref{sec:three}}.

Under the appropriate conditions (i.e. an incoming flux of ice and water), the above system is evolved until the surface temperature reaches freezing, i.e $\Tw(\hw(t^*),t^*)=T_\text{frz}=0^{\circ}C$. Here, $t^*$ denotes the critical time at which point freezing occurs at the corresponding water thickness, $\hw^*=\hw(t^*)$. Dependent on initial and boundary conditions, such a finite-time freezing event may not occur; in this work, we focus on situations where it does.

\subsection{Stage 2 of the three-layer model for ice-crystal icing} \label{sec:three}

Once a freezing event occurs at a critical time, $t = t^*$, and height, $\hw = \hw^*$, in the single-layer formulation of \S\ref{sec:water}, we proceed to the second stage in which three distinct water-ice-water layers are modelled. As is shown in the right side of figure~\ref{fig:scheme}, the domain is now bounded by $0 \leq z \leq \hw(t) + \hi(t) + \hsurf(t)$. This corresponds to internal water height $\hw$, middle ice layer $\hi$, and top surface water layer $\hsurf$. We therefore introduce the following notation for the absolute heights:
\begin{equation}
\begin{aligned} \label{eq:height_absolute}
\zwit(t) &\equiv \hw(t), \\
\ziwt(t) &\equiv \hw(t) + \hi(t), \\
\ztop(t) &\equiv \hw(t) + \hi(t) + \hsurf(t),
\end{aligned}
\end{equation}
in addition to the corresponding domain for each layer:
\begin{equation}
\begin{aligned}
\Omega_\text{b. water} \ \text{(bottom water region)} &= \{z: z\in [0, \zwit(t)]\}, \\
\Omega_\text{ice} \ \text{(middle ice region)} &= \{z: z\in [\zwit(t), \ziwt(t)]\}, \\
\Omega_\text{t. water} \ \text{(top water region)} &= \{z: z\in [\ziwt(t), \ztop(t)\}. \\
\end{aligned}
\end{equation}
We also need to solve for the temperatures within each of the three different layers, which are given by $\Tw(z,t)$ for $z \in \Omegabw$; $\Ti(z,t)$ for $z \in \Omegai$; and $T_\text{surf}(z,t)$ for $z \in \Omegatw$.

Only the temperatures of the internal water and ice layers will be solved for in this model. These are
\begin{subequations}\label{eq:DimBS2}
\begin{align}\label{eq:DimBS2tempwater}
\rho c_{\text{w}} \frac{\partial \Tw}{\partial t} &=k_{\text{w}}\frac{\partial^2 \Tw}{\partial z^2} && \text{for} \quad z \in \Omegabw, \\ \label{eq:DimBS2tempice}
\rho c_{\text{i}} \frac{\partial \Ti}{\partial t} &=k_{\text{i}}\frac{\partial^2 \Ti}{\partial z^2} && \text{for} \quad z \in \Omegai.
\end{align}
As noted in \S\ref{sec:EMM}, the above heat equations are typically solved under the quasi-steady assumption [cf. \cite{myers2001extension} for supercooled water and by \cite{bucknell2019three} for ice crystal icing]. Thus, the time-dependent left-hand sides are often neglected in implementations within the literature.

On the bottom substrate, we have 
\begin{equation}\label{eq:DimBS2heatflux}
\Tw(z,t)=\Tsubs \qquad \text{at $z=0$}. 
\end{equation}

In order to model the top water film, and in consideration of the fact that there is a complicated exchange of ice and water from the incomming flux, an additional assumption is made over the original Messinger model of \S\ref{sec:EMM}. Since the surface water film is typically very thin, \cite[p.~5]{bucknell2019icicle} assumes that the temperature gradient can be ignored and (since $T_{\text{surf}}(\ziwt(t),t)=0$), the top water film can be considered homogeneous in temperature:
\begin{equation}\label{eq:DimBS2tempsurf2}
T_{\text{surf}}(z,t) \equiv 0 \qquad \text{for} \quad z \in \Omegatw.
\end{equation}

We also provide the interfacial boundary conditions for the internal water and ice layers, which by assumption from \S\ref{sec:EMM} are all at the freezing temperature,
\begin{equation}
 \Tw(\zwit(t),t)=\Ti(\zwit(t),t)=\Ti(\ziwt(t),t) = 0.
 \label{eq:temperatureconstraint}
\end{equation}

The temperature gradient across the surface water layer is related to the heat flux on the exposed surface, hence by \eqref{eq:DimBS2tempsurf2} the net heat flux is zero. Then on the exposed surface,
\begin{equation}\label{eq:DimBS2substratetemp}
\Phi_\text{II}(\dot{m}_f) =0  \qquad \text{at $z=\ztop(t)$},
\end{equation}
where the flux, $\Phi_\text{II}(\mf)$, takes a similar form to that of $\Phi_{\text{I}}$ presented previously for the water-only case in \eqref{eq:Phi}: 
\begin{multline} \label{eq:Phi_II1}
\Phi_\text{II}(\mf) = \underbrace{h_{\text{tc}}(\Tsurf-T_{\text{rec}})}_{\text{convection}} + \underbrace{L_{\text{v}}\dot{m}_{\text{ev}}(\Tsurf)}_{\text{evaporation}}-\underbrace{L_{\text{f}}\mf}_{\text{melting/freezing}} \\
+\underbrace{\vphantom{\frac{1}{2}}c_{\text{w}}\dot{m}_{\text{imp}}(\Tsurf-T_\text{frz})}_{\text{sensible}}
-\underbrace{\frac{1}{2}\dot{m}_{\text{imp}}\bar{U}^2}_{\text{kinetic}},
\end{multline}
where from \eqref{eq:DimBS2tempsurf2},  $\Tsurf = 0 $, and as before $T_\text{frz}=0$ is the freezing temperature.

Studying \eqref{eq:Phi_II1}, we see that $\Phi_{II}$ has the same convection, evaporation, sensible,  and kinetic terms as the water-only case with $\Phi_I$, but now the surface temperature $T_{\text{surf}}$ has replaced the former $\Tw(\hw(t),t)$. Another difference is that the freezing/melting contributions on the right hand-side take a different form as we have entered Stage 2, and we no longer require that all particles melt. The melting contribution ($L_{\text{f}}\dot{m}_{\text{imp,i}}$) from $\Phi_\text{I}$ \eqref{eq:Phi} is then replaced with a freezing contribution, $-L_{\text{f}}\dot{m}_{\text{f}}$, in $\Phi_\text{II}$. 

In the end, \eqref{eq:DimBS2substratetemp} and \eqref{eq:Phi_II1} provide an equation which is solved for the mass flux between the ice and surface water layer which is freezing, denoted by $\mf$.

For the internal water growth of $\hw(t)$, a Stefan condition drives the water-ice interface, similar to the original extended Messinger model:
\begin{align}
   -\rho_{\text{w}}L_{\text{f}} \frac{\mathrm{d}\hw}{\mathrm{d}t}&= -k_{\text{w}} \frac{\partial \Tw}{\partial z} + k_{\text{i}} \frac{\partial \Ti}{\partial z} && \text{at} \quad z=\zwit(t).\label{eq:DimBS2water} \\ 
   \intertext{From \eqref{eq:temperatureconstraint}, the temperature within the ice layer should be invariant at the freezing temperature and the temperature gradient should be zero, hence $\Ti=0$ and ${\partial \Ti}/{\partial z} = 0$. The ice layer, $\hi(t)$, must consider the evolution of both interfaces, at $z=\zwit(t)$ and $z=\ziwt(t)$. It is modeled by}
    \rho_{\text{i}}\dd{\hi}{t} &=  - \rho_\text{w}\dd{\hw}{t} +(\dot{m}_{\text{imp,i}}+ \dot{m}_{\text{f}}). && ~ \label{eq:DimBS2ice} \\
    \intertext{Above, the first term on the right-hand side is from the Stefan condition, as any growth from the internal water layer corresponds to melting from the ice layer. The second term on the right-hand side, $\m_\text{imp,i}$, is from the impingement of penetrating ice particles. The last term on the right hand side, $\mf$, is the melting/freezing mass flux which is a solution of the boundary condition \eqref{eq:DimBS2heatflux}. 
    \endgraf 
    The surface water film, $h_{\text{surf}}(t)$, grows according to }
   \rho \frac{\mathrm{d}\hsurf}{\mathrm{d}t}&= \dot{m}_{\text{imp,w}} - \dot{m}_{\text{f}} - \dot{m}_{\text{ev}}(T_{\text{surf}}) && \text{at} \quad z=\ztop(t), \label{eq:DimBS2surfaceevolve}
\end{align}
where $\dot{m}_{\text{imp,w}}$ is the liquid portion of the incoming mass, and $\mf$ is the melting/freezing contribution between the surface water layer and the ice layer, as solved above from \eqref{eq:DimBS2heatflux}.

To close the system, we provide initial conditions for the heights of the different layers, which are given by
   \begin{equation}
        \hw(t^*) = \hw^*, \qquad  \hi(t^*) = 0.
    \end{equation}
\end{subequations}

\subsubsection*{A remark on the instantaneous passage of ice to the ice layer}

We mention one key remark with how the above governing equations are developed from \cite{bucknell2019three}. When considering the surface water film and ice layer, the authors allow the ice contributions from incoming particles to (instantaneously) penetrate through the water layer, so that ice is added directly to the ice layer below. This behaviour is implemented via a split of the impinging mass flux, $\mimp$, into the water ($\m_\text{imp,w}$) and ice ($\m_\text{imp,i}$) contributions. Thus, water is added to the top water-layer directly via a source term [cf. \eqref{eq:DimBS2surfaceevolve}] while ice is added directly to the ice layer [cf. \eqref{eq:DimBS2ice}].

\subsection{Non-dimensionalisation of the three-layer model} \label{sec:ff_nondim}

We begin by examining the different mass fluxes appearing in \eqref{eq:DimBS2ice} and \eqref{eq:DimBS2surfaceevolve}. We can split our impinging mass flux into water and ice contributions which is determined by the melt ratio, $M_r$. Thus, we introduce this parameter to distinguish between water and ice impingement, setting
\begin{equation} \label{eq:Mr}
\m_\text{imp,w} = \MR\mimp \qquad \text{and} \qquad \m_\text{imp,i} = (1-\MR) \mimp.
\end{equation}
We now scale our different mass flux by the total impingement:
\begin{equation}\label{eq:nonDimMass}
    \ndd{\m_\text{imp,w}} = \frac{ \m_\text{imp,w}}{\mimp}=\MR, \quad \ndd{\m_\text{imp,i}} = \frac{ \m_\text{imp,i}}{\mimp}=1-\MR, \quad \ndd{\mf} = \frac{ \mf}{\mimp}, \quad \ndd{\m_\text{ev}} = \frac{ \m_\text{ev}}{\mimp}.
\end{equation}

\subsubsection{Stage 1 (water only)} \label{sec:ff_nondim_S1}
We now nondimensionalise the water-only system of \eqref{eq:DimBS1}. The following non-dimensional scalings are introduced for the spatial and temporal variables: 
\begin{equation}
\begin{gathered}
\ndd{z}=\frac{{z}}{[H]}, \qquad \ndd{\hw}=\frac{{\hw}}{[H]}, \qquad \ndd{t}=\frac{t}{[t]}=\frac{\mimp t}{\rhow [H]}, \\
\ndd{\Tw}=\frac{{\Tw}}{T_{\text{rec}}}, \qquad \ndd{\Tsubs}=\frac{\Tsubs}{T_{\text{rec}}}.
\end{gathered}
\end{equation}
The time scale has been chosen to correspond to the rate at which mass enters the system via $\mimp$. Henceforth for the remainder of this section, we drop primes on the non-dimensional quantities. We also relabel the temperature $\Tw \mapsto T$ on the assumption that it is clear where the temperature is measured.

This yields the following non-dimensional governing equations: 
\begin{subequations}\label{eq:nonDimS1BVP}
\begin{align}
\Pe\frac{\partial {\nd{T}}}{\partial {\nd{t}}}&=\frac{\partial^2 {\nd{T}}}{\partial {\nd{z}}^2}, && \text{for $\nd{z}\in\nd{\Omegabw}$}, \label{eq:nonDimBS1temp} \\
\frac{\mathrm{d}{\nd{\hw}}}{\mathrm{d}{\nd{t}}}&= 1 - \nd{\dot{m}_{\text{ev}}}, && \text{at $\nd{z} = \nd{\hw}$}. \label{eq:nonDimBS1surfaceevolve} 
\end{align}
The temperature conditions \eqref{eq:DimBS2substratetemp} and \eqref{eq:DimBS2heatflux} at the solid substrate and free surface, respectively, satisfy
\begin{align}\label{eq:nonDimBS1substratetemp}
\nd{T} &=\nd{\Tsubs}, && \text{at $\nd{z} = 0$}, \\ \label{eq:nonDimBS1flux}
-\frac{\partial \nd{T}}{\partial \nd{z}} &= \nd{\Phi_\text{I}}(T(z, t)), && \text{at $\nd{z} = \nd{\hw}$},
\end{align}
Above, and from \eqref{eq:Phi} we have the flux, $\Phi_\text{I}$ defined as
\begin{equation} \label{eq:nonDimBS1Accretion}
 \nd{\Phi_\text{I}} \equiv  \Bi(\nd{T}-1) + \St\Lh\nd{\dot{m}_{\text{ev}}}(T) +\St (1-\MR) +\Pe\nd{T}-  \St\D.
\end{equation}  
\end{subequations}

We have also introduced the following nondimensional parameters: 
\begin{equation}\label{eq:nondimconstants}
\begin{gathered}
    \Pe = \frac{\mimp \cw [H]}{\kw}, \qquad \Bi=\frac{h_{\text{tc}}[H]}{k_{\text{w}}}, \qquad \St = \frac{\mimp L_\text{f}[H]}{\kw\Trec}, \\
    \D=\frac{\bar{U}^2}{2L_{\text{f}}}, \qquad \Lh = \frac{L_\text{v}}{L_\text{f}}, \qquad \Hc = \frac{\ci}{\cw},
\end{gathered}
\end{equation}
respectively corresponding to the P\'{e}clet number, Biot number, Stefan number, a ratio of kinetic to freezing energy, and the ratio of latent heats. 
We will discuss typical parameter ranges in \S\ref{sec:parameters}. For the water height, a reasonable estimate is $[H] \approx 10^{-4}$m \citep{bucknell2018ice}.

\subsubsection{Stage 2 (three-layer configuration) \label{sec:ff_nondim_S2}} 
In addition to the above, we nondimensionalise $\hi(t)$ and $\hsurf(t)$ with respect to $[H]$, as well as temperature with respect to $\Trec$. This yield the additional nondimensional quantities:
\begin{equation}
\ndd{\hi}=\frac{{\hi}}{[H]}, \quad \ndd{\hsurf}=\frac{{\hw}}{[H]}, \quad \ndd{\Ti} = \frac{\Ti}{\Trec}.
\end{equation}

It should be noted that in the three-layer model, the ice and top surface water layer are always assumed to be at $\Ti \equiv 0$ and $\Tsurf \equiv 0$. Therefore, only the temperature of the lower water layer must be solved:
\begin{subequations}
\begin{equation} \label{eq:nonDimBS2tempwater}
\Pe_\text{w} \frac{\partial \nd{T}}{\partial \nd{t}} =\frac{\partial^2 \nd{T}}{\partial \nd{z}^2},  \qquad \text{for} \quad \nd{z} \in \nd{\Omegabw}, 
\end{equation}
with $T = \Tw$ above. 

The layer growths are rewritten from \eqref{eq:DimBS2water}, \eqref{eq:DimBS2ice}, and \eqref{eq:DimBS2surfaceevolve} as
\begin{align}
\label{eq:nonDimBS2Stefan}
    \frac{\mathrm{d}\nd{\hw}}{\mathrm{d}\nd{t}}&=  \frac{1}{\St} \left(- \frac{\partial \nd{T}}{\partial \nd{z}} \right) &\text{at}& \quad z=\zwit(t), \\ \label{eq:nonDimBS2ice}
    \dd{\nd{\hi}}{\nd{t}} &= \frac{1}{\Ro} \bigg[ -\dd{\nd{\hw}}{\nd{t}}+(1-\MR)+ \nd{\mf} \bigg],  & & \\ \label{eq:nonDimBS2surfaceevolve}   
    \frac{\mathrm{d}\nd{\hsurf}}{\mathrm{d}\nd{t}}&= \MR - \nd{\mf} - \nd{\dot{m}_{\text{ev}}}(T = 0). 
\end{align}
\end{subequations}
Again, since $\Ti \equiv 0$, the ice temperature disappears in the Stefan condition of \eqref{eq:nonDimBS2Stefan}, and because $\Tsurf \equiv 0$, the surface water temperature is set to zero within the evaporative term of \eqref{eq:nonDimBS2surfaceevolve}. 

Above, we have introduced the additional conductivity and density ratios:
\begin{equation}
    \K = \frac{\ki}{\kw} \quad \text{and} \quad \Ro = \frac{\rhoi}{\rhow}.
\end{equation}
It remains to specify the boundary conditions on the temperature of the lower water layer. We have
\begin{subequations}
\begin{align}
\nd{T} &=\nd{\Tsubs} && \text{at $\nd{z} = 0$},\label{eq:3layersubstratecondition} \\ 
\nd{T}  &= 0 && \text{at $\nd{z} = \hw(t)$},\label{eq:3layertopwatercondition} \\ 
0 &= \nd{\Phi_\text{II}}(T = 0; \mf) && \text{at $\nd{z} = \nd{\hsurf}(t)$}, \label{eq:3layermfcondition}
\end{align}
\end{subequations}
The substrate boundary condition \eqref{eq:3layersubstratecondition} remains the same as the water-only case presented in \S{\ref{sec:ff_nondim_S1}}. At the surface, the temperature is assumed to be zero according to \eqref{eq:DimBS2tempsurf2}. Therefore, we consider the adjusted flux given previously by \eqref{eq:Phi_II1}. As before, by setting this flux to zero, the melting/freezing contribution, $\nd{\mf}$, is obtained as a solution from \eqref{eq:3layermfcondition}. 

It follows that in non-dimensional form,
\begin{equation} \label{eq:nonDimBS2Accretion}
 \nd{\Phi_\text{II}}(T = 0, \nd{\mf}) \equiv  -\Bi + \St\Lh\nd{\dot{m}_{\text{ev}}}(T = 0) - \St \D-\St \nd{\mf} = 0.
\end{equation}
Hence solving for $\nd{\mf}$ yields
\begin{equation}\label{eq:nonDimBS2Accretion2}
    \nd{\mf} \equiv \Lh\nd{\m_\text{ev}}(0)- \frac{\Bi}{\St} - \D.
\end{equation}
If $\nd{\mf} < 0$, this implies that melting rather than freezing occurs. The above expression can now be substituted into \eqref{eq:nonDimBS2ice} and \eqref{eq:nonDimBS2surfaceevolve}.

We remind the reader that the above set of equations are non-dimensional, where primes ($'$) have been dropped. In order to retrieve the dimensional forms, each quantity should be multiplied by its respective scaling, e.g. $z \mapsto [H]z$.

In summary, the solution of the three-layer model consists of solving: (i) three unknown heights, $\hw$, $\hi$ and $\hsurf$ using \eqref{eq:nonDimBS2Stefan} -- \eqref{eq:nonDimBS2surfaceevolve}; (ii) a temperature, $\Tw(z, t)$, for the water layer using \eqref{eq:nonDimBS2tempwater}; and (iii) a mass flux value $\mf$ for the melting/freezing between the top water and ice layer using the surface boundary condition \eqref{eq:nonDimBS2Accretion2}. In denoting $t=t^*$ as the time at which ice first forms and we transition to the three-layer model presented in this section, the initial conditions for $T(z,t^*)$ and $\hw(t^*)$ are that obtained from stage one in \S{\ref{sec:ff_nondim_S1}}, and additionally $\hi (t^*)=0$ and $\hsurf (t^*)=0$. Many parameters are required in this model, and these will be summarised and discussed in \S{\ref{sec:parameters}}.

\section{Formulation of the enthalpy model}\label{sec:dimensionalEnth}
The enthalpy model that we formulate in this section captures different physical phenomena to that of the three-layer model previously outlined in \S\ref{sec:fixedfront}. Rather than assuming that each of the water-ice-water phases are distinct layers separated by interfaces, we allow for the top ice and water layers to mix, forming a mixed-phase layer (denoted later by $\hm$). At high altitudes, the internal accretion can consist of partly melted ice particles \citep{Currie2012,currie2013altitude,Currie2014}, and therefore such mixed-phase models may better reflect the physical conditions encountered in ice-cystal icing. Below, we refer to this mixed phase layer as a \emph{mushy region}. Another advantage of the enthalpy formulation is its ability to capture phase changes;  for instance, the transition between an initial water phase and a mixed-phase solution due to surface ice accretion.

We begin by formulating the one-dimensional model with respect to dimensional quantities. The domain we consider is given by $0 \leq {t} < \infty$, and $0 \leq {z} \leq {\htotal}({t})$. Here, ${z}$ is the spatial coordinate orthogonal to the lower substrate, which lies at ${z}=0$. The unknown surface, denoted by ${z}={\htotal}({t})$, will form part of our solution. 

Following \cite{CrankBook1984}, the enthalpy, ${E}({z},{t})$, and Kirchoff transform on temperature, ${v}({z},{t})$, are introduced according to 
\begin{subequations}\label{eq:DimEnthalpy}
   \begin{align}\label{eq:DimEnthalpyE}
{E}(z,t) = \left\{
    \begin{aligned}
    & \rho c_{\text{w}}{T} + \rho L_{\text{f}} \qquad &\text{for~}{T}>0,\\
    & \in \left(0,\rho L_{\text{f}}\right] \qquad  &\text{for~} {T}=0,
    \end{aligned}
        \right.
\end{align}
\begin{equation}\label{eq:DimEnthalpyV}
 {v}(z,t)=k_{\text{w}}{T} \qquad \text{for~} {T} \geq 0,
\end{equation}
\end{subequations}
where ${T}({z},{t})$ is the usual temperature. We have assumed that the density of ice and water is the same, and given by the constant $\rho$. Further, $c_{w}$ is heat capacity of the fluid, $L_{\text{f}}$ is the latent heat of fusion, and $k_{\text{w}}$ is the thermal conductivity of the fluid. These constants are specified later in table~\ref{tab:physicalconditions}, in which typical parameter values are given.

We note that in most applications of the enthalpy formulation, an additional relation is specified in \eqref{eq:DimEnthalpyE} and \eqref{eq:DimEnthalpyV} for the case of ${T}<0$ \citep{CrankBook1984}, which describes ice. This relation must be included when examining all the different configurations of aircraft icing, as some can involve engine conditions or substrates which are below freezing temperature. However, in our current formulation, we shall only consider positive wet bulb temperature; then, the temperature only takes non-negative values. This assumption is consistent with experimental results reported by \cite{bartkus2018evaluation}, \cite{struk2018ice}, and \cite{bucknell2019icicle}. 

The enthalpy in \eqref{eq:DimEnthalpyE} is evolved according to the heat equation,
\begin{subequations}\label{eq:DimBVP}
\begin{equation}\label{eq:DimEnth}
\frac{\partial {E}}{\partial {t}}=\frac{\partial^2 {v}}{\partial {z}^2} \qquad \text{for} \quad 0 \leq {z} \leq {\htotal}({t}).
\end{equation}
On the solid substrate, we impose,
\begin{equation}\label{eq:Dimsubstratetemp}
{T}(0,{t})=\Tsubs,
\end{equation}
for the specified temperature $\Tsubs$. On the surface, $z = \htotal$, we also have a comparable flux condition to \eqref{eq:Phi} that comprises convection, kinetic, evaporation, and melting contributions: 
\begin{multline} \label{eq:DimAccretioncopy}
-k_{\text{w}}\frac{\partial {T}}{\partial {z}} = \Bigl[h_{\text{tc}}(T-T_{\text{rec}})\Bigr] 
+ \Bigl[L_{\text{v}}\dot{m}_{\text{ev}}(T)\Bigr]
+ \Bigl[L_{\text{f}}\dot{m}_{\text{imp,i}}\Bigr] \\
+ \Bigl[c_{\text{w}}\dot{m}_{\text{imp}}(T-T_\text{frz})\Bigr]
- \Bigl[\frac{1}{2}\dot{m}_{\text{imp}}\bar{U}^2\Bigr].
\end{multline}

We wish to use an alternative form of the sensible and melting heat fluxes (third and fourth square brackets), which is more convenient for the enthalpy formulation. Recalling the definition of enthalpy for temperatures above freezing \eqref{eq:DimEnth}, and evaluating our temperature at the accretion surface, we have $E(h,t) = \rho \cw T(h,t) + \rho L_f$. In addition, we define our impinging enthalpy by $\Eimp =  \MR \rho L_f$. Thus, we can write the difference between our enthalpy at the surface and impinging enthalpy as, 
\begin{equation*}
\frac{\mimp}{\rho}( E(h,t) - \Eimp) =\vphantom{\frac{\mimp}{\rho}} \mimp \cw T(h,t) + \m_\text{imp,i} L_f.
\end{equation*}
Above, we have used the fact that $1-\MR = \m_\text{imp,i}/\mimp$, which follows from \eqref{eq:Mr}. The right hand-side includes the sensible and melting heat contributions, as written in \eqref{eq:Phi}, while the left hand-side expresses the enthalpy-specific version. Substituting the above into \eqref{eq:DimAccretioncopy} we have our enthalpic surface boundary condition given by
\begin{equation}\label{eq:DimAccretion}
-k_{\text{w}}\frac{\partial {T}}{\partial {z}} = \Bigl[\underbrace{\vphantom{\frac{\mimp}{\rho}}h_{\text{tc}}({T}-T_{\text{rec}}}_{\textrm{convection}})\Bigr] + \Bigl[\underbrace{\vphantom{\frac{\mimp}{\rho}} L_{\text{v}}\dot{m}_{\text{ev}}({T})}_\textrm{evaporation}\Bigr]+\Bigl[\underbrace{\frac{\mimp}{\rho}(E-\Eimp)}_{\textrm{freeze/melt+sensible}}\Bigr]-\Bigl[\underbrace{\vphantom{\frac{\mimp}{\rho}}\frac{1}{2}\dot{m}_{\text{imp}}\bar{U}^2}_\textrm{kinetic}\Bigr]
\end{equation}
at $z=h_{\text{total}}(t)$.

Additionally, the unknown total height, $\htotal(t)$, evolves according to
\begin{equation}\label{eq:Dimsurfaceevolve}
\rho \frac{\mathrm{d}{\htotal}}{\mathrm{d}{t}}= \dot{m}_{\text{imp}} - \dot{m}_{\text{ev}}(T \rvert_{z=\htotal(t)}),
\end{equation}
\end{subequations}
where $\dot{m}_{\text{imp}}$ is the constant impingement flux, and $\dot{m}_{\text{ev}}({T})$ is the evaporation rate. As reviewed in Appendix~\ref{sec:evap}, the evaporation must be temperature-dependent, and by using the model proposed by \cite{wexler1983thermodynamic} it is assumed to take the form
\begin{equation}\label{eq:Dimmdotevformula}
    \nd{\m_\text{ev}}(T) = A\left[ \mathrm{e}^{c_1T^{-1}+c_2+c_3T+c_4T^2+c_5T^3+c_6 \log (T)} - P_{\text{vap,sat},\infty}\right].
\end{equation}
Here, $A$, $P_{\text{vap,sat},\infty}$, and $c_i$ are dimensional constants specified in Appendix~\ref{sec:evap}. We define the constant $A$ in equation \eqref{eq:Appmevformula}, and experimentally determined values for $c_i$, from \cite{wexler1983thermodynamic}, are given in table~\ref{tab:vapourcoeffs}.

\subsection{Nondimensionalisation \label{sec:Enth_nondim}}

We now nondimensionalise the boundary-value problem \eqref{eq:DimBVP} for the enthalpy formulation, in which ${E}$ and ${v}$ are related to the temperature, ${T}$, by \eqref{eq:DimEnthalpyE} and \eqref{eq:DimEnthalpyV}, respectively. We nondimensionalise $z$ and $\htotal$ with respect to the length scale $[H]$, and $t$ with the time scale $[t]=\rho [H]/\mimp$. Additionally, the temperature, $T$, is nondimensionalised  with respect to the recovery temperature, $\Trec$, $v$ with respect to $\kw \Trec$, and the enthalpy, $E$, by $\rho \cw \Trec$.
Combined, this yields the relations
\begin{equation}\label{eq:enthnondimscaling}
\begin{aligned}
&z'=\frac{{z}}{[H]}, \qquad \htotal'=\frac{{\htotal}}{[H]}, \qquad t'=\frac{{t}\mimp}{\rho [H]},\\ 
&T'=\frac{{T}}{T_{\text{rec}}}, \qquad v'=\frac{{v}}{k_{\text{w}}T_{\text{rec}}}, \qquad E'=\frac{{E}}{\rho c_{\text{w}} T_{\text{rec}}},
\end{aligned}
\end{equation}
in which non-dimensional quantities are denoted with primes. The equations governing these nondimensional quantities are derived next, in which we abuse notation by removing primes from the each of the quantities defined in \eqref{eq:enthnondimscaling}. 

\begin{figure}
    \centering
    \includegraphics{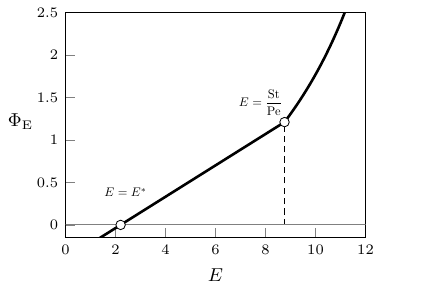}
    \caption{Plot of $E$ vs $\Phi_\textrm{E}$  as described in \S\ref{sec:Enth_nondim}. Parameters are nondimensionalised given by \cref{tab:nd_param} ($\St=1.618$, $\Pe=0.185$, $T_{\text{subs}}=10$, $\Bi=0.070$, $L=6.711$, $\MR = 0.2$, and $D=0.028$).}
    \label{fig:PhiEvsE}
\end{figure}

Firstly, we consider the equations relating $E(z,t)$ and $v(z,t)$ to the temperature, $T(z,t)$, from \eqref{eq:DimEnthalpyE} and \eqref{eq:DimEnthalpyV}, which yields in nondimensional form
\begin{subequations}\label{eq:nonDimBVP}
\begin{equation} \label{eq:vE}
\nd{E}(z, t) = 
 \left\{
    \begin{aligned}
    & \nd{T} +  \frac{\St}{\Pe}   &\text{for~}\nd{T}>0,\\
    &\in \left(0,  \frac{\St}{\Pe}  \right]  &\text{for~} \nd{T}=0,
    \end{aligned} \right.
\qquad \text{and} \qquad  v(z,t)=T ~~~ \text{for}~~~ T \geq 0.
\end{equation}
Here, $\Pe$ is the P\'{e}clet number, and $\St$ is the Stefan number, both defined previously in equation \eqref{eq:nondimconstants}.
Next, the nondimensional heat equation for the enthalpy is found by substituting relations \eqref{eq:enthnondimscaling} into \eqref{eq:DimEnth}, which yields
\begin{equation}\label{eq:nonDimEnth}
\Pe\frac{\partial {\nd{E}}}{\partial {\nd{t}}}=\frac{\partial^2 {\nd{v}}}{\partial {\nd{z}}^2} \qquad \text{for} \quad 0 \leq {\nd{z}} \leq {\nd{\htotal}}({\nd{t}}).
\end{equation}
The two boundary conditions for the temperature are found from \eqref{eq:Dimsubstratetemp} and \eqref{eq:DimAccretion} to be
\begin{equation}
{\nd{T}}(0,t) =\Tsubs, \label{eq:nonDimsubstratetemp}
\end{equation}
\begin{equation}
-\frac{\partial \nd{T}}{\partial \nd{z}} =  \Phi_E \equiv \Bi(\nd{T}-1) +\St\Lh\nd{\dot{m}_{\text{ev}}}(\nd{T}) +\Pe(\nd{E}-\nd{E}_{\text{imp}})- \St \D \qquad \text{at $z=\htotal(t)$} \label{eq:nonDimAccretion},
\end{equation}
where we show the variation of $\Phi_\text{E}(E)$ with regards to $E$ in the right side of \cref{fig:PhiEvsE}. In \cref{fig:PhiEvsE} we note the value that the heat flux is zero, denoted by $\Estarp$, which we will discuss later in \S{\ref{sec:Enth_interpret}}. Note also that in \eqref{eq:nonDimAccretion}, we have defined $E_{\text{imp}} = \MR\St/\Pe$. 

The evolution equation for the interface is given by
\begin{equation}\label{eq:nonDimsurfaceevolve}
\frac{\mathrm{d}{\nd{\htotal}}}{\mathrm{d}{\nd{t}}}= 1 - \nd{\dot{m}_{\text{ev}}}({T}\rvert_{{z}={\htotal}(t)}).
\end{equation}
\end{subequations} 

In the above, $\m_{\text{ev}}(T)$ is the nondimensional evaporative mass flux. Our expression for this, given later in \eqref{eq:Hyland}, is found by substituting for the nondimensional relations \eqref{eq:nonDimMass} and \eqref{eq:enthnondimscaling} into the dimensional evaporative mass flux in \eqref{eq:Dimmdotevformula}. This expression contains several experimental fitted constants, which we specify in Appendix \ref{sec:evap}.

Our nondimensional parameters are the same as those defined in equation \eqref{eq:nondimconstants} of \S\ref{sec:ff_nondim}, in which the three-layer formulation was presented. Note that in the lower boundary condition \eqref{eq:nonDimsubstratetemp}, we have defined $T_{\text{subs}}^{\prime}=T_{\text{subs}}/T_{\text{rec}}$, which is nondimensional, and then removed the prime notation.

In summary, the solution of the enthalpy model consists of solving for the height, $\htotal(t)$, and the enthalpy, $E(z,t)$. These solutions are coupled via the boundary condition \eqref{eq:nonDimAccretion} and evolution equation \eqref{eq:nonDimsurfaceevolve}. As an initial condition, at $t=0$ we will specify $h_{\text{total}}(0)=0$ and $T(0,0)=T_{\text{subs}}$.

\subsection{Interpretation and prediction of $\hw$ and $\hm$ from the enthalpy solution} \label{sec:Enth_interpret}

The three-layer model from \S\ref{sec:fixedfront} provides explicit solutions for the height of each layer corresponding to the lower water layer, $\hw(t)$, the middle ice layer, $\hi(t)$, and the top water layer, $\hsurf(t)$. However, the enthalpy method only yields the total height, 
\begin{equation}
\htotal(t) = \hw(t) + \hm(t),
\end{equation}
and does not explicitly provide insight on the lower water layer, $\hw$, and upper mixed-phase layer, $\hm$. In this section, we discuss how these components can be extracted from a computed solution, and how to measure the proportion of ice in the mixed-phase layer. This then facilitates comparison with the traditional three-layer model.

\begin{figure}
    \centering
    \includegraphics[scale=1]{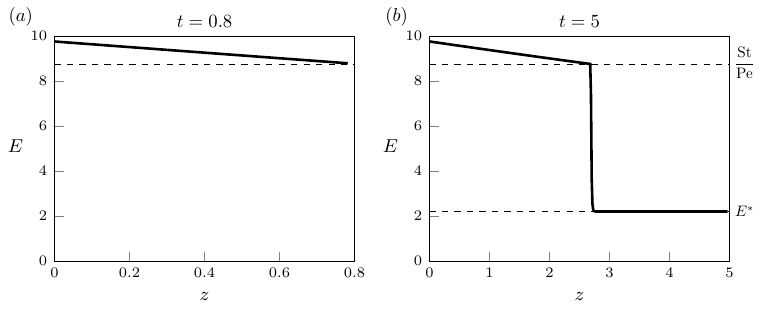}
    \caption{A typical solution profile for the Enthalpy, $E(z,t)$, is shown at $t=0.8$ in $(a)$ and $t=5$ in $(b)$. The solution in $(a)$, with $E>\St/\Pe$ corresponds to a pure water layer. The solution in $(b)$ contains both a pure water layer for $0 \leq z < 2.7$, and a mixed-phase region for $2.7 \leq z \leq h_{\text{total}}$. In $(b)$, there is a thin transition region about $z=2.7$ in which the proportion of ice particles in the mixed-phase layer varies from 0\% to 75\%. This numerical simulation used the method outlined in Appendix~{\ref{sec:flag_Enth}}, with parameter values $\St=1.618$, $\Pe=0.185$, $T_{\text{subs}}=10$, $\Bi=0.070$, $L=6.711$, $\MR = 0.2$, and $D=0.028$. A value of $N=400$ grid points was used for simulations.} 
    \label{fig:Enth_interpret}
\end{figure} 

\begin{enumerate}[label={(\roman*)},leftmargin=*, align = left, labelsep=\parindent, topsep=3pt, itemsep=2pt,itemindent=0pt ]
\item We first anticipate the numerical computations shown in \cref{sec:num} and show a typical solution, $E(z, t)$, in figure~\ref{fig:Enth_interpret}$(a)$ corresponding to $t = 0.8$. The simulation begins with the initial condition of $\htotal = 0$ and $T =\Tsubs>0$, and therefore the entire domain will initially consist of water only. This remains the case as long as $E(\htotal,t) > \St/\Pe$ across the spatial domain, which by relation \eqref{eq:vE} is equivalent to $T >0$. In this case, we have that all the accumulation is water, and therefore $\hw(t)=\htotal(t)$ and $\hm(t)=0$. This stage lasts until a mixed-phase region forms with $T(z,t)=0$.

\item After the inception of the mixed-phase layer induced by ice accretion, the domain consists of a lower water layer, within which $E(z,t) \geq \St/\Pe$, and the mixed-phase layer with $\St/\Pe > E(z,t)>0$. The boundary, $z=\hw(t)$, between these two regions is defined by $E(\hw(t),t)=\St/\Pe$. Then the mixed-phase layer height is subsequently given by $\hm(t)=\htotal(t)-\hw(t)$. An example solution in this regime is shown in figure~\ref{fig:Enth_interpret}$(b)$, corresponding to $t =5$.

\item Note that the enthalpy of the mixed-phase layer may be further restricted to the range $\St/\Pe > E(z,t)  \geq \Estarp$, where $\Estarp$ is the ``balancing enthalpy". The constant $\Estarp$ may be found analytically by equating the right-hand side of \eqref{eq:nonDimAccretion} to zero. It is further assumed that $\nd{T}=0$, and subbing in for the impinging enthalpy discussed earlier in \S{\ref{sec:dimensionalEnth}}, we find 
\begin{equation} \label{eq:Estar}
\Estarp \equiv (\MR\St -\St \Lh\nd{\m}_{\text{ev}} + \Bi + \St\D)/\Pe. 
\end{equation} 

As we are currently not considering the sub-freezing regime, $\Estarp \in (0,\St/\Pe]$ and its value depends on each of the nondimensional constants in \eqref{eq:Estar}. For instance, for those parameters given in figure~\ref{fig:Enth_interpret}, $E^* \approx 2.2$. 
\end{enumerate}

Through examination of the enthalpy within the mixed-phase layer, we can also extricate the ice contribution.
Dividing by the enthalpy jump, $\St/\Pe$, gives the mushy phase enthalpy fraction
\begin{equation} \label{eq:beta}
    \beta = \frac{\Estarp}{\St/\Pe} =  \MR  + (\Bi/\St + \D - \Lh\nd{\m}_{\text{ev}}),
\end{equation}
which is analogous to the melt ratio, $\MR$, but relates to the accretion instead of the impingement. Similarly to dividing the total impingement into water and ice contributions, via the melt ratio as in \eqref{eq:nonDimMass}, we can estimate the total water and ice contributions by 
\begin{equation} \label{eq:enth_ice}
   \text{water component} = \beta \hm, \qquad \text{ice component} = (1-\beta) \hm. 
\end{equation}
Note that the enthalpy fraction $\beta$ can be related to the freezing contribution of the three layer model, calculated earlier in \eqref{eq:nonDimBS2Accretion2} via
\begin{equation}
    \beta = \MR - \nd{\mf}.
\end{equation}

It is useful to compare our definitions of $M_r$ and $\beta$ with parameters introduced in the melt/freeze-dominated regime model of \cite{bartkus2018evaluation,Bartkus2019}. With $m_0$ and $n_0$ defined in their eqn (7) and (10) of \cite{bartkus2018evaluation}, we have $m_0 = (\beta-\MR)/(1-\MR)$ and $n_0 = (\MR-\beta)/\MR$, 
which correspond to the fraction of ice that melts and fraction of water that freezes, respectively. In the situation of $\beta = 1$, we have running wet conditions $m_0 = 1$, as the balancing enthalpy is at the pure water enthalpy value and all impinging ice is melting. If $\beta = 0$, there is no water contribution in the enthalpy as $n_0 = 1$, and all impinging water is freezing.

\section{Parameters of the model} \label{sec:parameters}

One of the many challenges in the study of ice crystal icing is the number of parameters involved, and the identification of the different icing stages in the system \citep{mason2020engine}. Although we focus primarily on the accretion dynamics in this work, full models may also consider the effects of \emph{e.g.} ice-particle tracking, and resultant impact and shedding on the surfaces. Consideration of these extra effects will introduce additional parameters.

We consider parameters as roughly categorised into four categories. 

The first group of parameters corresponds to well-established physical properties, such as the properties of water, ice and air \citep{myers2001extension}. 

The second group of parameters is related to icing conditions in aircraft engines, and are documented in various experiments. These include quantities such as Mach number, wet bulb temperature, total temperature, substrate temperature, substrate heat flux, melt ratio of particles, total pressure, particle diameter distribution size, and so on \citep{Currie2012,hauk2016theoretical,Baumert2018Experimental, bucknell2018experimental,Malik2024}. Note that some of these are known in the context of experiments and do not necessarily translate fully to realistic engine conditions.

\afterpage{ 
\begin{table}
    \centering
    \begin{tabular}{ccccc}
         & \textsc{name}  & \textsc{unit} & \textsc{Magnitude} & \textsc{Reference}\\
        $\rho$ & Fluid density & $\text{kg}/\text{m}^3$  & 1000 & \cite{myers2001extension}\\
        $\rhoi$ & Ice density & $\text{kg} / \text{m}^{3}$  & 917 & \cite{myers2001extension}\\
          $c_{\text{w}}$ & Heat capacity of fluid  & $\text{J}/(\text{kg} \cdot \text{K})$ & 4218 & \cite{myers2001extension}\\
          $c_{\text{i}}$ & Heat capacity of ice  & $\text{J} /( \text{kg} \cdot \text{K})$ & 2050 & \cite{myers2001extension}\\
          $c_{\text{a}}$ & Heat capacity of air  & $\text{J} /( \text{kg} \cdot \text{K})$ & 1014 & \cite{myers2001extension}\\
        $h_\text{tc}$ & Heat transfer coefficient & $\text{J} /( \text{s} \cdot \text{m}^{2}\cdot \text{K}) $ & 50--650 & \cite{villedieu2014glaciated, currie2020physics} \\
          $k_{\text{w}}$ & Fluid thermal conductivity & $\text{J}/( \text{s} \cdot \text{m} \cdot \text{K})$  & 0.571 & \cite{myers2001extension}\\
          $k_{\text{i}}$ & Ice thermal conductivity & $\text{J}/( \text{s}\cdot \text{m} \cdot \text{K})$  & 2.18 & \cite{myers2001extension}\\
        $L_{\text{f}}$ & Latent heat of fusion  & $\text{kJ} / \text{kg}$ & 334 & \cite{myers2001extension}\\
          $L_{\text{v}}$ & Latent heat of vaporisation  & $\text{kJ} / \text{kg}$ & 2200--2500 & \cite{myers2001extension,currie2020physics}\\
          $M_\text{a}$ & Molar mass of air & $\text{g} / \text{mol}$ & 29 & \cite{meija2016atomic}\\
          $M_\text{w}$ & Molar mass of water & $\text{g} / \text{mol}$ & 18 & \cite{meija2016atomic}\\
          $\Tsubs$ & Substrate temperature & K  & 273.15--293.15 & \cite{bucknell2018ice, Malik2024}\\
          $\Trec$ & Recovery temperature & K  & 273.15--293.15 & \cite{struk2015ice,bucknell2018ice}\\
          $\dot{m}_\text{imp}$ & Impingement flux & $\text{kg} / (\text{s} \cdot \text{m}^{2})$ & 0.15--1.08 & \cite{currie2011fundamental,bucknell2018experimental}\\
          $\bar{U}$ & Velocity & $\text{m} / \text{s}$ & 40-225 & \cite{struk2015ice,bucknell2018ice}  \\ 
          $Le$ & Lewis number & [-] & 0.8--1.2 & \cite{Currie2012}\\ 
          $\MR$ & Melt Ratio  &  [-] & 0.05--0.25 & \cite{struk2015ice,bucknell2018ice}\\ 
          $b$ & Lewis number exponent & [-] & 0.33 or 0.65 & \cite{bucknell2019three}
     \end{tabular}
    \caption{Typical values of dimensional quantities under engine representative conditions, as used in the existing literature. Note that some estimates given in the references are more specified and single-valued. Others are variable under variable engine or flow conditions.\label{tab:physicalconditions} }    
\end{table}
}

The third group of parameters correspond to the physical set-up; these are often associated with calculation of the airflow via computational fluid dynamics. This relates to the particle tracking and heat transfer, and parameters such as the heat transfer coefficient are needed to then relate to other quantities such as the mass transfer coefficient. These are correlated by the nondimensional parameters such as the Sherwood, Nusselt, Prandtl numbers, etc. \citep{bucknell2018ice}. In the full system, parameters such as the pressure and shear stresses will be critical in determining the effects of water runback and break-off \citep{mason2020engine}.  

Finally, consider parameters which are used in most modern codes, which may be empirical or phenomenological, and that may be employed to fill a gap in existing physics-based understanding. These include parameters such as the collection/sticking efficiency \citep{currie2013altitude} or parameters related to other physical effects such as erosion or shedding, which have either been taken from experiments \citep{bartkus2018evaluation} or relate to numerous other parameters such as the erosion efficiency in \cite{trontin2016comprehensive, bucknell2019icicle}.

A complete listing of dimensional parameters used in this work are summarised in \cref{tab:physicalconditions}. In \cref{tab:nd_param}, we list the resultant nondimensional parameters, derived using values from \cref{tab:physicalconditions}. The non-dimensional values in this table establish the ``baseline values" which are referred-to in the rest of the paper.

\subsection{Thresholds for freezing} \label{sec:freeze_threshold}
Earlier, at the end of \S{\ref{sec:water} we mentioned that a finite time freezing event might not occur if the surface temperature never reaches freezing. Here we obtain critical thresholds in parameters whereby we would not expect ice growth in either the three-layer or enthalpy models. By examining \eqref{eq:nonDimBS1Accretion} (or \eqref{eq:nonDimAccretion}), we set $\Phi_I=0$ and identify the temperature at which this occurs: 
\begin{equation}
    T = \frac{\Bi-\St(\Lh\mev(T)+1-\MR-D)}{\Bi+\Pe}.
\end{equation}

Note that above, $\mev(T)$ is a nonlinear function of temperature. We are typically interested in the case of $T>0$ and $T \rightarrow 0^+$ for the critical temperature. Thus we can develop an approximate threshold:
\begin{equation} \label{eq:Bi_crit}
   \Bi = \Bi_\text{crit}(\St, L, \mev, M_r, D) \equiv \St(\Lh\mev(0)+1-\MR-D),
\end{equation}
with freezing expected for $\Bi < \Bi_\text{crit}$. Therefore we observe an approximate  linear relationship between the Biot and Stefan numbers which categorises the region of freezing. 

Note that as either the evaporation increases, the melt ratio decreases, or the kinetic ratio is reduced, the gradient of the critical curve increases, thus increasing the potential for freezing. Conversely, with increasing melt ratio or kinetic ratio or decreasing evaporative flux, this reduces the slope, resulting in a higher threshold for freezing. These relationships are visualised in \cref{fig:Freeze_threshold}.

\begin{figure}
    \centering
    \includegraphics[scale=1]{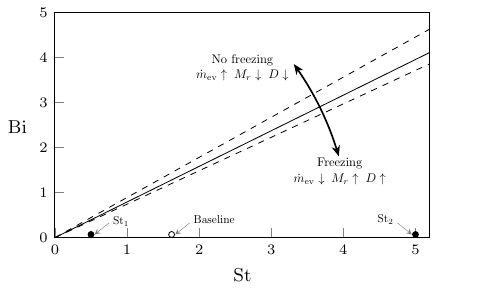}
    \caption{The critical curve separating freezing from non-freezing conditions as measured via \eqref{eq:Bi_crit}. The solid line corresponds to the baseline conditions given in \cref{tab:nd_param} discussed in \S{\ref{sec:parameters}}. The upper dashed line modifies $\MR = 0.1$ from baseline, thus decreasing the melt ratio. The lower dashed line modifies $RH = 0.65$ from baseline, thus increasing the relative humidity, and decreasing the evaporative flux. The critical lines rotate anticlockwise for increasing evaporative flux, decreasing melt ratio, or decreasing kinetic ratio. The black and white circles denote parameter values used in \cref{fig:Pevststar}, which consist of our baseline Stefan number, along with two other values, denoted $\St_1$ and $\St_2$ respectively.}
    \label{fig:Freeze_threshold}
\end{figure}

\afterpage{
\begin{table}
    \centering
    \begin{tabular}{cccc}
    \textsc{Symbol}  & \textsc{Meaning} & \textsc{Form} & \textsc{Value used} \\
          $\Pe$ & P\'{e}clet number & $\mimp \cw [H]/\kw$  & 0.185 \\
          $\St$ & Stefan number  & $ \mimp L_\text{f}[H]/\kw\Trec$  & 1.618\\
         $\Bi$ & Biot number & $h_{\text{tc}}[H]/k_{\text{w}}$ &  0.070 \\        
         $\D$ & Ratio of kinetic to phase change energies & $\bar{U}^2/2L_{\text{f}}$ & 0.028 \\
         $\MR$ & Melt Ratio & ${\dot{m}}_{\text{imp,w}}/\mimp$ & 0.2 \\
          $\Hc$ & Ratio of heat capacities & $\ci/\cw$  &  0.486 \\
          $\Lh$ & Ratio of latent heats  & $L_\text{v}/L_\text{f}$  & 6.711 \\
          $\K$ & Ratio of thermal conductivity & $\ki/\kw$ & 3.680\\
          $\Ro$ & Ratio of densities & $\rhoi/\rhow$ & 0.917\\
          $RH$ & Relative humidity & [--] & 0.45 \\
           $\beta$ & Enthalpy fraction & $\Estarp/(\St/\Pe)$ & $0.252^*$
    \end{tabular}
    \caption{Nondimensional parameters in the ice crystal icing models and their typical values used in this work. *Note that $\beta$ is found as a solution based on the other parameter values according to \eqref{eq:beta}. \label{tab:nd_param} } 
\end{table}
}

\section{Asymptotic analysis for small P\'{e}clet number} \label{sec:asympt}
The P\'{e}clet number, $\Pe$, is important in determining the governing form of heat transfer. In the case that the P\'{e}clet number is small, we have the effect of diffusion dominating over advection in the transfer of heat. From examining the definition of the P\'{e}clet number given in \cref{tab:nd_param}, $\Pe=\mimp \cw [H]/\kw$, we can determine characteristic values of the P\'{e}clet number. In our baseline case, we consider $\mimp=0.25kg/(m^2s)$ and $[H] = 10^{-4}m$ [cf. \cite{bucknell2018ice,currie2020physics,villedieu2014glaciated}]; these we combine with the typical values of $\cw$ and $\kw$ given in \cref{tab:physicalconditions} to calculate a P\'{e}clet value of 0.185. In different conditions, such as those where the impinging mass flux is smaller, we can have $\Pe \approx 10^{-2}$. Conversely, if dealing with a larger mass flux, or hotter engine surface temperatures which produce a thicker water layer, we could have $\Pe \approx 10$.

The situation of small P\'{e}clet number leads to many simplifications when studying the system. The heat equation becomes quasi-static which means that the temperature will have a linear profile and thus, easier implementation in icing code.
In this section, we shall develop leading-order asymptotics of the enthalpy model in the limit of $\Pe \to 0$ and the first-order corrections.

\subsection{Water-only state ($0 \leq t < t^*$)} \label{sec:0th_water}
In this section we provide leading- and first-order asymptotic solutions for the running wet conditions, in the limit of $\Pe \to 0$. We first present this for the enthalpy formulation, but it can be verified that the analysis of the water-only regime in \S{\ref{sec:ff_nondim_S1}} follows identically, which we expand on later in this section. 

We expand the temperature, $T$, and water height, $h$, as asymptotic expansions in powers of small P\'{e}clet number as
\begin{subequations}\label{eq:asymptoticexpansionPe}
\refstepcounter{equation}\label{eq:asymptoticexpansionPea}
\refstepcounter{equation}\label{eq:asymptoticexpansionPeb}
\begin{equation}
    T(z, t) \sim T_0(z,t)+\Pe T_1(z,t)+\cdots \quad \text{and} \quad 
    h(t) \sim h_0(t)+\Pe h_1(t)+\cdots.
    \tag{\ref*{eq:asymptoticexpansionPea},$b$}
\end{equation}
\end{subequations}
Note that from the enthalpy relations \eqref{eq:nonDimBVP}, we have $v(z,t) \sim T_0 + \Pe T_1 +\cdots$ and $\Pe E(z,t) \sim \St +\Pe T_0 + \Pe^2 T_1 + \cdots$. The governing equations for the leading- and first-order components of each expansion \eqref{eq:asymptoticexpansionPe} are now derived. We begin by expanding the heat equation \eqref{eq:nonDimEnth}, which yields at $O(\Pe^0)$ and $O(\Pe)$ the equations
\begin{subequations}\label{eq:asymptoticheatequationPe}
\refstepcounter{equation}\label{eq:asymptoticheatequationPea}
\refstepcounter{equation}\label{eq:asymptoticheatequationPeb}
\begin{equation}
\frac{\partial^2 T_0}{\partial z^2}=0 \quad \text{and} \quad \frac{\partial^2 T_1}{\partial z^2}=\frac{\partial T_0}{\partial t},
\tag{\ref*{eq:asymptoticheatequationPea},$b$}
\end{equation}
\end{subequations}
respectively. These are second-order PDEs, for which one boundary condition arises from the fixed substrate temperature \eqref{eq:nonDimsubstratetemp}, which when expanded yields the conditions
\begin{subequations}\label{eq:substrateBCPe}
\refstepcounter{equation}\label{eq:substrateBCPea}
\refstepcounter{equation}\label{eq:substrateBCPeb}
\begin{equation}
T_0(0,t)=\Tsubs \quad \text{and} \quad T_1(0,t)=0.
\tag{\ref*{eq:substrateBCPea},$b$}
\end{equation}
\end{subequations}
The second boundary condition for temperature comes from the surface flux condition \eqref{eq:nonDimAccretion}, which we now expand as $\Pe \to 0$. Note that this boundary condition contains components evaluated at the surface $z=h(t)$, which itself is expanded asymptotically in powers of $\Pe$ by \eqref{eq:asymptoticexpansionPeb}. Taylor expanding each of these components then allows for the isolation of terms of $O(\Pe^{0})$ and $O(\Pe)$, which yields
\begin{subequations}\label{eq:fluxBCPe}
\begin{align}
\label{eq:fluxBCPea}
-\frac{\partial T_0}{\partial z}&=\Bi\big[T_0-1\big] +\St L \mev(T_0)+\St(1-\MR - D),\\
\label{eq:fluxBCPeb}
 - \frac{\partial T_1}{\partial z} -h_1 \frac{\partial^2 T_0}{\partial z^2}&=\bigg(\Bi+\St L \frac{\mathrm{d} \mev}{\mathrm{d}T}(T_0) \bigg)\bigg(T_1 +h_1 \frac{\partial T_0}{\partial z}\bigg)+T_0,
\end{align}
\end{subequations}
both of which hold at $z=h_0(t)$. The sensible components of the heat flux are not retained at leading-order, which is consistent with \cite{roychowdhury2023ice} who observed that the sensible heat from the particle temperature was negligible compared with other heat sources. Note the dependence on $h_0$ and $h_1$ in conditions \eqref{eq:fluxBCPe} above. The equations governing these may be derived from the evolution equation \eqref{eq:nonDimsurfaceevolve}. As before, we use expansions \eqref{eq:asymptoticexpansionPe} to Taylor expand $T(h,t) \sim T_0(h_0,t)+\Pe [T_1(h_0,t)+h_1T_{0z}(h_0,t)] +\cdots$, and collect terms at each order of $\Pe$ in \eqref{eq:nonDimsurfaceevolve}. This yields at $z=h_0(t)$ the evolution equations
\begin{subequations}\label{eq:hevolutionPe}
\refstepcounter{equation}\label{eq:hevolutionPea}
\refstepcounter{equation}\label{eq:hevolutionPeb}
\begin{equation}
\frac{\mathrm{d}h_0}{\mathrm{d}t}=1- \mev(T_0) \quad \text{and} \quad \frac{\mathrm{d}h_1}{\mathrm{d}t}=-\bigg(h_1 \frac{\partial T_0}{\partial z} +T_1\bigg)\frac{\mathrm{d} \mev}{\mathrm{d}T}(T_0),
\tag{\ref*{eq:hevolutionPea},$b$}
\end{equation}
\end{subequations}
with initial conditions $h_0(0)=0$ and $h_1(0)=0$.

These equations form a closed system for $T_0$, $h_0$, $T_1$, and $h_1$. For instance, $T_0$ satisfies the 
 PDE \eqref{eq:asymptoticheatequationPea}, with boundary conditions \eqref{eq:substrateBCPea} and \eqref{eq:fluxBCPea}. The evolution equation \eqref{eq:hevolutionPea} and initial condition $h_0(0)=0$ then govern $h_0$. Note however the coupling between these; the equations for $T_0$ involve $h_0$ and vice versa. Similar equations are found for the $O(\Pe)$ solutions of each expansion, but with additional forcing terms involving leading-order solutions.

The solutions of PDEs \eqref{eq:asymptoticheatequationPe} that satisfy boundary conditions \eqref{eq:substrateBCPe} at $z=0$ are given by
 \begin{equation}\label{eq:asymptoticsolutionsTempPe}
T_0(z,t) =a_0(t)z +\Tsubs  \quad \text{and} \quad T_1(z,t)= a_1(t)z +\frac{a_0^{\prime}(t)z^3}{6},
\end{equation}
where $a_0(t)$ and $a_1(t)$ are to be determined and prime $(^\prime)$ denotes differentiation. Substitution of solutions \eqref{eq:asymptoticsolutionsTempPe} into equations \eqref{eq:fluxBCPe} and \eqref{eq:hevolutionPe} then yields the four equations
\begin{subequations}\label{eq:bcwithTsolPe}
\begin{align}\label{eq:bcwithTsolPea}
\frac{\mathrm{d}h_0}{\mathrm{d}t}&=1-\mev\big(a_0h_0+\Tsubs \big),\\
\label{eq:bcwithTsolPeb}
 \frac{\mathrm{d}h_1}{\mathrm{d}t} &= -\bigg(a_0h_1+a_1h_0+\frac{a_0^{\prime}h_0^3}{6} \bigg)\frac{\mathrm{d}\mev}{\mathrm{d}T}\big(a_0h_0+\Tsubs\big),\\
 \label{eq:bcwithTsolPec}
 a_0(t)&=\Bi \big[1-a_0h_0-\Tsubs\big]-\St L \mev \big( a_0h_0+\Tsubs \big)+\St (D+\MR-1),\\
 \label{eq:bcwithTsolPed}
 &\hspace{-8mm}\begin{aligned}
 a_1(t)=&-\frac{a_0^{\prime}h_0^2}{2}-\bigg(\Bi+\St L \frac{\mathrm{d}\mev}{\mathrm{d}T} \big( a_0h_0+\Tsubs\big) \bigg) \bigg(a_0h_1+a_1h_0 +\frac{a_0^{\prime}h_0^3}{6}\bigg)\\
 &-(a_0h_0 + \Tsubs),
 \end{aligned}
\end{align}
\end{subequations}
with solutions given by $h_0(t)$, $h_1(t)$, $a_0(t)$, and $a_1(t)$. The difficulty now in solving system \eqref{eq:bcwithTsolPe} above is the form of the function $\mev(T)$. For our numerical solutions calculated later in \S\ref{sec:num}, we use the nonlinear model \eqref{eq:Dimmdotevformula} from \cite{wexler1983thermodynamic} in which $\mev(T)=A\exp{(c_1/T + c_2 + c_3 T + c_4 T^2 +c_5 T^3 + c_6 \log{(T)})}-AP]$. In \S\ref{sec:simplifymevconst}, we assume this evaporation rate is constant to make analytical progress.

Note that an equivalent asymptotic analysis of the three-layer model from \S\ref{sec:ff_nondim_S1} leads to the same results as those presented in this section. This is because when only the initial water layer is present, the difference between this and the enthalpy formulation is in the form of the surface flux boundary conditions, \eqref{eq:nonDimBS1flux} and \eqref{eq:nonDimAccretion}. However, by the definition of $E_{\text{imp}} = \MR\St/\Pe $, we have that these two conditions are equivalent.

\subsubsection{An approximation for the evaporation, $\mev$}\label{sec:simplifymevconst}
We now simplify the form of $\mev(T)$ in order to make analytical progress in finding the solution of equations \eqref{eq:bcwithTsolPe}.
Simplifications used in numerical work by previous authors include linear and piece-wise linear approximations for this function \citep{myers2001extension,bucknell2019three}. However, if we take $\mev(T)=\alpha_1 +\alpha_2 T$ for instance, the exact solutions of \eqref{eq:bcwithTsolPe} are given in terms of special functions which are difficult to evaluate directly. 

In the remainder of this section, we consider a constant approximation for this function in order to obtain more explicit solutions. Our approximation here, $\mev{(T)}=\mev{(0)}$, assumes that the evaporation rate is unaffected by temperature. Under this assumption, equations \eqref{eq:bcwithTsolPe} have the solutions
\begin{equation}\label{eq:asympconstexsols}
\left.\begin{aligned}
h_0(t)&=[1-\mev(0)]t, && a_0(t)=\frac{\Bi [1-\Tsubs]+\St[D+\MR-1-L\mev(0)] }{1+\Bi h_0},\\
h_1(t)&=0, && a_1(t)=-\frac{a_0h_0 +\Tsubs}{1+\Bi h_0}-\frac{a_0^{\prime} h_0^2[3+\Bi h_0]}{6[1+\Bi h_0]}.
\end{aligned} \right\}
\end{equation}
We will now use these solutions to calculate the time, and height of the water layer, when freezing first occurs.

\subsection{The freezing point, $t=t^*$}
The time, $t^*$, at which freezing temperature is reached at the water surface, $z=h(t)$, may be found by solving the equation $T(h(t^*),t^*)=0$. In expanding both $T(z,t)$ and $h(t)$ as $\Pe \to 0$ as in \eqref{eq:asymptoticexpansionPe}, and also expanding 
\begin{equation}
t^* \sim t^*_0 + \Pe t^*_1 + \cdots,
\end{equation}
we have that $t_0^*$ is a solution of the equation $T_0(h_0(t_0^*),t_0^*)=0$, and $t_1^*$ is a solution of $T_1(h_0(t_0^*),t_0^*) + t_1^*\partial_t T_0(h_0(t_0^*),t_0^*)+[h_1(t_0^*) + t_1^*h_0^{\prime}(t_0^*)]\partial_z T_0(h_0(t_0^*),t_0^*)=0$. Substitution of the
solutions for $T_0$ and $T_1$ given in \eqref{eq:asymptoticsolutionsTempPe} then yields
\begin{equation}\label{eq:leadingfreezingeq}
a_0(t_0^*) h_0(t_0^*) +\Tsubs=0,
\end{equation}
from which $t_0^*$ is a solution, and an explicit solution for $t_1^*$ given by
\begin{equation}\label{eq:firstordertimefre}
t_1^*=-\frac{a_1(t_0^*)h_0(t_0^*)+a_0(t_0^*) h_1(t_0^*) + a_0^{\prime}(t_0^*) h_0^3(t_0^*)/6}{a_0^{\prime}(t_0^*)h_0(t_0^*)+a_0(t_0^*)h_0^{\prime}(t_0^*)}.
\end{equation}

We may now calculate $t_0^*$ by substituting solutions \eqref{eq:asympconstexsols} for $h_0$ and $a_0$ into equation \eqref{eq:leadingfreezingeq}. The height of the leading-order water layer at this critical time may then also be found. Combined, these yield the solutions
\begin{subequations}\label{eq:crit}
\begin{align}\label{eq:critt0}
 t_0^* &= \frac{\Tsubs}{(1-\mev(0))(\St[1+L \mev(0)-D-\MR]-\Bi)},\\\label{eq:crith0}
  h_0(t_0^*) &= \frac{\Tsubs}{\St[1+L \mev(0)-D-\MR]-\Bi}.  
\end{align}
\end{subequations}
Note that these are the leading-order approximations as $\Pe \to 0$ for the freezing time and water height. The fact that we have found the first-order correction to $h(t)$ to be identically zero in \eqref{eq:asympconstexsols} is a consequence of assumption of constant evaporative flux made only in this section. The first-order correction, $t_1^*$, to the freezing time can be calculated from \eqref{eq:firstordertimefre}. This leads to 
\begin{equation}
    t_1^*=-\frac{a_1(t_0^*)t_0^* + a_0^{\prime}(t_0^*) [1-\mev(0)]^2(t_0^*)^3/6}{a_0^{\prime}(t_0^*)t_0^*+a_0(t_0^*)}.
\end{equation}
Later in \S\ref{sec:num}, we use the first two orders calculated in this section, $t^* \sim t_0^* + \Pe t_1^*$, to compare with values obtained numerically.

\subsection{Three-layer state for $t>t^*$} \label{sec:0th_3}
As noted in \S\ref{sec:three}, a key assumption of the three-layer model is that the ice layer and the top water layer are fixed at freezing temperature. Thus, we previously set $  \Ti(z, t) = 0$ and $\Tsurf(z, t)=0$ in their respective regions.

We begin by expanding the heat equation \eqref{eq:nonDimBS2tempwater}, which yields at $O(\Pe^0)$ and $O(\Pe)$ the equations
\begin{subequations}\label{eq:asymptoticheatequationPe2}
\refstepcounter{equation}\label{eq:asymptoticheatequationPea2}
\refstepcounter{equation}\label{eq:asymptoticheatequationPeb2}
\begin{equation}
\frac{\partial^2 T_0}{\partial z^2}=0 \quad \text{and} \quad \frac{\partial^2 T_1}{\partial z^2}=\frac{\partial T_0}{\partial t},
\tag{\ref*{eq:asymptoticheatequationPea2},$b$}
\end{equation}
\end{subequations}
respectively. These are second-order PDEs, for which one boundary condition arises from the fixed substrate temperature \eqref{eq:3layersubstratecondition}, for which expansion yields
\begin{subequations}\label{eq:substrateBCPe2}
\refstepcounter{equation}\label{eq:substrateBCPea2}
\refstepcounter{equation}\label{eq:substrateBCPeb2}
\begin{equation}
T_0(0,t)=\Tsubs \quad \text{and} \quad T_1(0,t)=0.
\tag{\ref*{eq:substrateBCPea2},$b$}
\end{equation}
\end{subequations}
The other boundary condition for temperature arises from fixing the water-ice interface to be at freezing temperature, $T(\hw(t),t) = 0$. Expanding both $T$ and $\hw$ as $\Pe \to 0$ yields
\begin{subequations}\label{eq:interfaceBCPe2}
\refstepcounter{equation}\label{eq:interfaceBCPea2}
\refstepcounter{equation}\label{eq:interfaceBCPeb2}
\begin{equation}
 T_0(h_0,t)=0 \quad \text{and} \quad T_1(h_0,t)+h_1(t)T_{0z}(h_0,t)=0.
\tag{\ref*{eq:interfaceBCPea2},$b$}
\end{equation}
\end{subequations}
Next, we consider the evolution equations for the widths of the lower water layer, $\hw(t)$, ice layer, $\hi(t)$, and top water layer, $\hsurf(t)$. Beginning with equation \eqref{eq:nonDimBS2Stefan} for $\hw \sim h_0+\Pe h_1 +\cdots$ yields 
\begin{subequations}\label{eq:nonDimBS2Stefan2}
\refstepcounter{equation}\label{eq:nonDimBS2Stefan2a}
\refstepcounter{equation}\label{eq:nonDimBS2Stefan2b}
\begin{equation}
   \frac{\mathrm{d}\nd{h_0}}{\mathrm{d}\nd{t}}=  -\frac{1}{\St}\frac{\partial \nd{T}_0}{\partial \nd{z}} \quad \text{and} \quad
    \dd{h_1}{t} = -\frac{1}{\St} \biggl(\frac{\partial \nd{T}_1}{\partial \nd{z}} + h_1\frac{\partial^2 T_0}{ \partial z^2} \biggr),
\tag{\ref*{eq:nonDimBS2Stefan2a},$b$}
\end{equation}
\end{subequations}
at $z=h_0(t)$.
Expanding as well the ice layer height, $\hi \sim h_\text{ice}^{(0)} + \Pe h_\text{ice}^{(1)} + \cdots$, and top water layer $ \hsurf \sim \hsurf^{(0)} + \Pe \hsurf^{(0)}$, we have from equations \eqref{eq:nonDimBS2ice} and \eqref{eq:nonDimBS2surfaceevolve}
\begin{equation}\label{eq:nonDimBS2ice2}
\left.\begin{aligned}
  \dd{\nd{h_\text{ice}^{(0)}}}{\nd{t}} &= \frac{1}{\Ro} \bigg[ -\dd{\nd{h_0}}{\nd{t}}+(1-\MR)+ \nd{\mf} \bigg], \qquad   && \dd{\nd{h_\text{ice}^{(1)}}}{\nd{t}} = -\frac{1}{\Ro}\dd{\nd{h_1}}{\nd{t}},\\
    \frac{\mathrm{d}\hsurf^{(0)}}{\mathrm{d}\nd{t}}&= \MR - \nd{\mf} - \nd{\dot{m}_{\text{ev}}}(0), \qquad && \frac{\mathrm{d} \hsurf^{(1)}}{\mathrm{d}t}=0.
\end{aligned} \right\}
\end{equation}

The initial conditions for the three-layer model are given at $t=t^*$. Those for the ice and top water-layer are $\hi(t^*)=0$ and $\hsurf(t^*)=0$. For the lower water height and temperature, we have $\hw(t^*)=\tilde{h}_{\text{water}}(t^*)$ and $T(z,t^*)=\tilde{T}(z,t^*)$, where $\tilde{h}_{\text{water}}(t^*)$ and $\tilde{T}(z,t^*)$ are the functions obtained at the onset of freezing in the single layer model. Expanding each of these as $\Pe \to 0$, along with $t^*\sim t_0^*+\Pe t_1^* +\cdots$ yields the initial conditions
\begin{equation}\label{eq:ICsfor3layerPe}
\left.\begin{aligned}
T_0(z,t_0^*)&=\tilde{T}_0(z,t_0^*), \qquad && T_1(z,t_0^*)=\tilde{T}_1(z,t_0^*)+t_1^* [\partial_{t}\tilde{T}_0(z,t_0^*)-\partial_{t}T_0(z,t_0^*)],\\
h_0(t_0^*)&=\tilde{h}_0(t_0^*) \equiv h_0^*, \qquad && h_1(t_0^*)=\tilde{h}_1(t_0^*)+t_1^*[\tilde{h}^{\prime}_0(t_0^*)-h^{\prime}_0(t_0^*)],\\
h_\text{ice}^{(0)} (t_0^*)&=0, \qquad && h_\text{ice}^{(1)} (t_0^*)=-t_1^* h_\text{ice}^{(0)\prime}(t_0^*),\\
\hsurf^{(0)} (t_0^*)&=0, \qquad && \hsurf^{(1)} (t_0^*)=-t_1^* \hsurf^{(0)\prime}(t_0^*).
\end{aligned} \right\}
\end{equation}

\subsubsection{Leading-order solutions}
We begin by solving PDE \eqref{eq:asymptoticheatequationPea2} for the leading-order temperature, $T_0$, which with boundary conditions \eqref{eq:substrateBCPea2} and \eqref{eq:interfaceBCPea2} yields
\begin{equation}\label{eq:T0_leading}
T_0(z,t) = \Tsubs \bigg(1-\frac{z}{h_0(t)}\bigg).
\end{equation}
The leading-order temperature profile \eqref{eq:T0_leading} can then be substituted into \eqref{eq:nonDimBS2Stefan2a}, which yields a nonlinear ODE for $h_0(t)$. By integrating this and applying the initial condition from \eqref{eq:ICsfor3layerPe}, we obtain the solution
\begin{equation}
     h_0(t) = \sqrt{(h_0^*)^2+\frac{2\Tsubs}{\St}(t-t_0^*)}.
\end{equation}
The leading-order solutions for the ice and surface water layer, $h_\text{ice}^{(0)}$ and $\hsurf^{(0)}$, can be found by integrating \eqref{eq:nonDimBS2ice2} with initial conditions from \eqref{eq:ICsfor3layerPe}, yielding 
\begin{align}
    h_\text{ice}^{(0)}(t) &= \frac{1}{\Ro} \bigg[h_0^*-\sqrt{(h_0^*)^2+\frac{2\Tsubs}{\St}(t-t_0^*)}  +(1-\MR+ \nd{\mf})(t-t_0^*) \bigg],\\
    \hsurf^{(0)}(t) &= \big(\MR - \nd{\mf} - \nd{\dot{m}_{\text{ev}}}(0)\big)(t-t_0^*),
\end{align}
respectively.

\subsubsection{First-order correction}
We begin by solving for the first-order temperature profile, $T_1$. Substitution of $T_0$ from
\eqref{eq:T0_leading} into the governing PDE \eqref{eq:asymptoticheatequationPeb2} yields a second-order problem with a forcing term, which we integrate along with boundary conditions \eqref{eq:substrateBCPeb2} and \eqref{eq:interfaceBCPeb2} to find
\begin{equation}\label{eq:firstordertempts2}
    T_1(z,t)= \Tsubs\bigg[ \frac{z^3}{6h_0^2}\frac{\mathrm{d}h_0}{\mathrm{d} t} + \bigg(\frac{h_1}{h_0^2}-\frac{1}{6}\frac{\mathrm{d}h_0}{\mathrm{d}t}\bigg)z\bigg].
\end{equation}
The ODE for $h_1(t)$ is found by substituting \eqref{eq:firstordertempts2} into \eqref{eq:nonDimBS2Stefan2b}, which gives
\begin{equation}
\dd{h_1}{t} + \frac{\Tsubs}{\St h_0^2}h_1 = - \frac{\Tsubs}{3\St}\dd{h_0}{t}.
\end{equation}
This linear first-order ODE may be solved using an integrating factor, with the initial condition from \eqref{eq:ICsfor3layerPe}, to find
\begin{equation}
     h_1(t) = -\frac{\Tsubs^2}{3\St^2}\frac{t-t_0^*}{h_0(t)}+\left(1-\mev(0)-\frac{\Tsubs}{\St h_0^*}\right)\frac{t_1^* h_0^*}{h_0(t)}.
\end{equation}
To solve for $h_\text{ice}^{(1)}$ and $\hsurf^{(1)}$, we integrate the respective equation from \eqref{eq:nonDimBS2ice2} and enforce the initial conditions from \eqref{eq:ICsfor3layerPe}. This yields
\begin{align}
     h_\text{ice}^{(1)}(t) &= \frac{1}{\Ro}\bigg[\frac{\Tsubs^2}{3\St^2}\frac{t-t_0^*}{h_0(t)}-\left(1-\mev(0)-\frac{\Tsubs}{\St h_0^*}\right)\frac{t_1^* h_0^*}{h_0(t)} +\left(\MR-\mev(0)-\mf \right)t_1^*\bigg], \\
     \hsurf^{(1)} &= -t_1^*\big(\MR - \nd{\mf} - \nd{\dot{m}_{\text{ev}}}(0)\big).
\end{align}

\subsubsection{Summary of asymptotic results for $t>t^*$}
We summarise the leading and first order results. The temperature profiles are given by
\begin{subequations} \label{eq:3Layer_sol}
\begin{align}  \label{eq:T0_sol}
    T_0(z,t) &= \Tsubs \bigg(1-\frac{z}{h_0(t)}\bigg),\\ \label{eq:T1_sol}
    T_1(z,t) &= \Tsubs^2\frac{z^3}{{h_0^3(t)}}+\left(\frac{\Tsubs h_1(t)}{h_0(t)}-\frac{\Tsubs^2}{6\St}\right)\frac{z}{h_0(t)}.
    \end{align}
The height is the water film is given by    
    \begin{align} \label{eq:h0_sol}
    h_0(t) &= \sqrt{(h_0^*)^2+\frac{2\Tsubs}{\St}(t-t_0^*)},\\ \label{eq:h1_sol}
    h_1(t) &= -\frac{\Tsubs^2}{3\St^2}\frac{t-t_0^*}{h_0(t)}+\left(1-\mev(0)-\frac{\Tsubs}{\St h_0^*}\right)\frac{t_1^* h_0^*}{h_0(t)}.
    \end{align}
    The ice thickness was solved to yield
    \begin{align}\label{eq:B0_sol}
    h_\text{ice}^{(0)}(t) &= \frac{1}{\Ro} \bigg[h_0^*-\sqrt{(h_0^*)^2+\frac{2\Tsubs}{\St}(t-t_0^*)}  +(1-\MR+ \nd{\mf})(t-t_0^*) \bigg],\\ \label{eq:B1_sol}
    h_\text{ice}^{(1)}(t) &= \frac{1}{\Ro}\bigg[\frac{\Tsubs^2}{3\St^2}\frac{t-t_0^*}{h_0(t)}-\left(1-\mev(0)-\frac{\Tsubs}{\St h_0^*}\right)\frac{t_1^* h_0^*}{h_0(t)} +\left(\MR-\mev(0)-\mf \right)t_1^*\bigg].
    \end{align}
    Finally, the surface water film evolves as
    \begin{align} \label{eq:hs0_sol}
    \hsurf^{(0)} &= (\MR - \nd{\mf} - \nd{\dot{m}_{\text{ev}}}(0))(t-t_0^*),\\ \label{eq:hs1_sol}
    \hsurf^{(1)} &= -t_1^*(\MR - \nd{\mf} - \nd{\dot{m}_{\text{ev}}}(0)).
\end{align}
\end{subequations} 

\subsection{Enthalpy mixed-phase layer for $t > t^*$} \label{sec:0th_enth}
As discussed in \S\ref{sec:dimensionalEnth}, for $t>t^*$, a mush layer appears above the water layer, and there is a sudden drop of enthalpy from $\St/\Pe$ to a value between $0$ and $\St/\Pe$. 
Again we examine the $\Pe \to 0$ limit and conclude that in the region $z \in [0, \hw]$, $\partial_{zz} v = 0$, and therefore $v^{(0)}(z) = a(t)z + b(t) $. Note in addition that we can verify that the temperature in the mushy region is everywhere zero, hence $v = 0$ for $z \in [\hw, \htotal]$. it follows

\begin{equation}
    v^{(0)}(z,t) = \begin{cases}
        a_1(t) z + b_1(t), & \text{ for } 0 \leq z \leq \hw, \\
        0, & \text{ for } \hw < z \leq \htotal.
        \end{cases}
\end{equation}
Note as well that $b_1(t) = \Tsubs$ by the wall condition. We also require a matching condition imposing continuity of temperature at $\hw$. This yields $v^{(0)}=0$ at $z=\hw$ and in turn $a_1(t) = -\Tsubs/\hw$. Note that the water layer height $\hw$ will be different with the value in the three-layer model at $t>t^*$ as the stefan condition is changed, which is discussed below. There is a change in gradient $\partial_z v$ at $z=\hw$ because $v \equiv 0$ for $z > \hw$.

In order to close the system, it remains to determine the water-line position $\hw$. 
For our enthalpy model, we do not have a full phase transition between water and ice, but rather a transition between water and a mush. This idea was shown earlier in \cref{fig:Enth_interpret}, where the mush will settle at some intermediate enthalpy $0 \leq E^* \leq \St/\Pe$. This reduces the latent heat jump required for phase change, where now it is now the difference between the enthalpy value of water, $\St/\Pe$, and the enthalpy of the mush, $\Estarp$. Thus, the effective enthalpy jump is given by $\St/\Pe\left( 1-\beta \right)$ where $\beta$, as in equation \ref{eq:beta}, is the fraction of water in the mushy region. The change of water layer height can be determined using
\begin{equation} \label{eq:EnthStefan}
    \dd{\hw}{t} = -\frac{1}{\St_{\text{eff}}} \pd{v}{z}\Bigr|_{z = \hw^-}, 
\end{equation}
where we have defined an effective Stefan number by
\begin{equation}
    \St_{\text{eff}} \equiv \St\left(1- \beta\right). 
\end{equation}
The mush layer can be determined by the difference between the total growth and the water layer. We can now write the governing equations similar to \S{\ref{sec:0th_3}}. Beginning with \eqref{eq:EnthStefan} for $\hw\sim h_0 + \Pe h_1 + \cdots$ we can write
\begin{subequations}\label{eq:nonDimEnthStefan}
\refstepcounter{equation}\label{eq:nonDimEnthStefan2a}
\refstepcounter{equation}\label{eq:nonDimEnthStefan2b}
\begin{equation}
   \frac{\mathrm{d}\nd{h_0}}{\mathrm{d}\nd{t}}=  -\frac{1}{\Steff}\frac{\partial \nd{T}_0}{\partial \nd{z}} \quad \text{and} \quad
    \dd{h_1}{t} = -\frac{1}{\Steff} \biggl(\frac{\partial \nd{T}_1}{\partial \nd{z}} + h_1\frac{\partial^2 T_0}{ \partial z^2} \biggr),
\tag{\ref*{eq:nonDimEnthStefan2a},$b$}
\end{equation}
\end{subequations}
at $z=h_0(t)$. Expanding the mush layer $\hm \sim h_\text{mush}^{(0)} + \Pe h_\text{mush}^{(1)}$ we get
\begin{subequations}\label{eq:nonDimmush}
\refstepcounter{equation}\label{eq:nonDimmusha}
\refstepcounter{equation}\label{eq:nonDimmushb}
\begin{equation}
   \frac{\mathrm{d}\nd{h_\text{mush}^{(0)}}}{\mathrm{d}\nd{t}}= \dd{\htotal}{t} - \dd{h_0}{t} \quad \text{and} \quad
    \dd{h_\text{mush}^{(1)}}{t} = -\dd{h_1}{t}.
\tag{\ref*{eq:nonDimmusha},$b$}
\end{equation}
\end{subequations}
Similar to the previous section, we can write the initial conditions of both layers as
\begin{equation}
\left.\begin{aligned}
 &h_0(t_0^*)=\tilde{h}_0(t_0^*) \equiv h_0^*, \qquad && h_1(t_0^*)=\tilde{h}_1(t_0^*)+t_1^*[\tilde{h}^{\prime}_0(t_0^*)-h^{\prime}_0(t_0^*)],\\
 &h_\text{mush}^{(0)} (t_0^*)=0, \qquad && h_\text{mush}^{(1)} (t_0^*)=-t_1^* h_\text{mush}^{(0)\prime}(t_0^*).
\end{aligned} \right\}
\end{equation}

The enthalpy system can be solved nearly identically to \S{\ref{sec:0th_3}} to obtain the following results. The temperature profile is given by
\begin{subequations} \label{eq:enthalpy_sol}
\begin{align} \label{eq:Te0_sol}
    T_0(z,t) &= \Tsubs\bigg(1-\frac{z}{h_0(t)}\bigg),\\ \label{eq:Te1_sol}
    T_1(z,t) &= \Tsubs^2\frac{z^3}{{h_0(t)}^3}+\left(\frac{\Tsubs h_1(t)}{h_0(t)}-\frac{\Tsubs^2}{6\Steff}\right)\frac{z}{h_0(t)}.
    \end{align}
    The internal water grows as
    \begin{align}\label{eq:he0_sol}
    h_0(t) &= \sqrt{(h_0^*)^2+\frac{2\Tsubs}{\St_{\text{eff}}}(t-t_0^*)},\\ \label{eq:he1_sol}
    h_1(t) &= -\frac{\Tsubs^2}{3\St_{\text{eff}}^2}\frac{t-t_0^*}{h_0(t)}+\left(1-\mev(0)-\frac{\Tsubs}{\St_{\text{eff}} h_0^*}\right)\frac{t_1^* h_0^*}{h_0(t)},
    \end{align}
    and the mixed-phase region is solved to give
    \begin{align}\label{eq:hm0_sol}
    h_\text{mush}^{(0)}(t) &= h_0^*-\sqrt{(h_0^*)^2+\frac{2\Tsubs}{\St_{\text{eff}}}(t-t_0^*)}  +(1-\mev)(t-t_0^*) ],\\ \label{eq:hm1_sol}
    h_\text{mush}^{(1)}(t) &= \frac{\Tsubs^2}{3\St_{\text{eff}}^2}\frac{t-t_0^*}{h_0(t)}-\left(1-\mev(0)-\frac{\Tsubs}{\St_{\text{eff}} h_0^*}\right)\frac{t_1^* h_0^*}{h_0(t)}, 
\end{align}
\end{subequations}
which are solutions valid for $t > t^*$.

\section{Numerical simulations} \label{sec:num}

In this section, we present typical numerical simulations showing comparisons between the three-layer formulation and the new enthalpy-based formulation. 

\subsection{Numerical details for the three-layer formulation}

The simulation is initiated in Stage 1 (\S\ref{sec:ff_nondim_S1}) with only the water film present. A change of variables, with $\xi = h/\hw \in [0,1]$, is used to account for the growing domain. The profile height, $\hw$, is evolved using a forward Euler scheme applied to \eqref{eq:nonDimBS1surfaceevolve}, while the temperature, given by \eqref{eq:nonDimBS1temp}, is solved using an implicit finite difference scheme. Typically, it is sufficient to discretise the spatial domain using $N = 100$ points. 

Once the temperature reaches freezing, we enter Stage 2 (\S {\ref{sec:ff_nondim_S2}}). We then solve for the additional two profile heights, $\hi(t)$ and $\hsurf(t)$. The bottom water temperature is solved via  \eqref{eq:nonDimBS2tempwater}, but now with a top condition setting the temperature to zero \eqref{eq:temperatureconstraint}. Once the surface temperature is at freezing, the evaporation mass flux will remain constant, and thus $\nd{\mf}$ from \eqref{eq:nonDimBS2Accretion2} will also constant. The three heights are all solved using explicit Euler time stepping applied to \eqref{eq:nonDimBS2Stefan}--\eqref{eq:nonDimBS2surfaceevolve}. For the simulations presented in this section, we use time steps of $\Delta t = 0.01$ for \cref{fig:evolve_three_enth} and \cref{fig:evolve_three_enth2}, and $\Delta t = 0.0001$ for \cref{fig:Pe_vs_t5} and \cref{fig:Pevststar}.

\subsection{Numerical details for the enthalpy formulation \label{sec:num_Enth}}
For the enthalpy formulation presented in \S{\ref{sec:Enth_nondim}}, in order to solve the enthalpy PDE \eqref{eq:nonDimBVP}, we rescale the problem and implement an implicit scheme via a flag-update method. This idea of a `flag update' is similar to the one implemented by \cite{bridge2007mixture}. After nondimensionalising the problem, we can rewrite our enthalpy by \eqref{eq:vE} in terms of the temperature and latent heat contribution. 

We perform a change of variables, with 
\begin{equation}
    \xi = \frac{z}{\htotal(t)},
\end{equation}
so that the spatial domain is fixed $\xi \in [0, 1]$. Then $\bar{E}(z,t) \rightarrow E(\xi, t), \; \bar{v}(z,t) \rightarrow v(\xi, t), \; \bar{T}(z,t) \rightarrow T(\xi, t)$. For example, the PDE is now 
\begin{equation}
    \Pe \left[ \pd{E}{t} - \frac{\xi}{\htotal}\frac{\mathrm{d}\htotal}{\mathrm{d}t}\pd{E}{\xi}   \right]
    = \frac{1}{\htotal^2}\frac{\partial^2 v}{\partial \xi^2}.
    \label{eq:Enth_PDE_coord}
\end{equation}
We discretise the spatial domain with $N$ points, 
\begin{equation}
\xi_i = \frac{i-1}{N-1}, \qquad i = 1, \ldots, N.
\end{equation}
Let $\mathbf{e}$ be the solution vector. Then at time $t = t_n$, we have
\begin{equation} \label{eq:sol_vector}
    \mathbf{e}(t_n) = [T_{1}^n, \; \hdots,\; T_{N}^n, \hdots, \;  S_{1}^n, \; \hdots,\; S_{N}^n]^T,
\end{equation}
where $T_i^n$ and $S_i^n\in [0, \St/\Pe]$ are the respective temperature and enthalpy and at $(\xi_i, t_n)$.

We then seek to solve the system
\begin{equation} \label{eq:num_system}
    \mathbf{Me}^{n+1} = \mathbf{f}^n,
\end{equation}
where $\mathbf{M}$ is a  $2 N\times 2 N$ sparse matrix. Above, the vector $\mathbf{f}$ is split into two components. The first half of the entries corresponds to the enthalpy at each point in the domain at the previous time step. The second half provides the latent heat contribution, as will be given by a regime flag update. Overall, it can be written as 
\begin{equation}
    \mathbf{f} = [E_{1}, \; \hdots, \; E_{N},\; 
     \phi_{1}, \; \hdots, \; \phi_{N}]^T,
\end{equation}
where $\phi_{i} = 0 \textrm{ or } \phi_{i}=\St/\Pe$, the value of which depends on the flag [cf. \cref{sec:flag_Enth}] at the $i$th grid point.

For intuition purposes, the matrix $\mathbf{M}$ can be decomposed into four smaller matrices, each of size $N \times N$,
\begin{equation} \label{eq:matrix_M}
    \mathbf{M} = \begin{pmatrix}
  \mathbf{T} & \mathbf{S}\\ 
  \mathbf{A} & \mathbf{\hi}
\end{pmatrix}.
\end{equation}
The matrices $\bold{T}$ and $\bold{S}$ are each tridiagonal matrices constructed by the finite difference approximation applied to equation \eqref{eq:Enth_PDE_coord}. The entries of matrices $\bold{A}$ and $\bold{B}$ are either 0 or 1 on the main diagonal and are dependent on the flag value at each grid-point, which connects to the latent heat jump. Further details of the construction of these matrices and the flag updates are given in Appendix~{\ref{sec:flag_Enth}}. 

\subsection{Numerical results} \label{sec:num_results}

\subsubsection{Dynamics of the water/ice layers} \label{sec:num_layers}

\begin{figure} 
    \centering
    \includegraphics{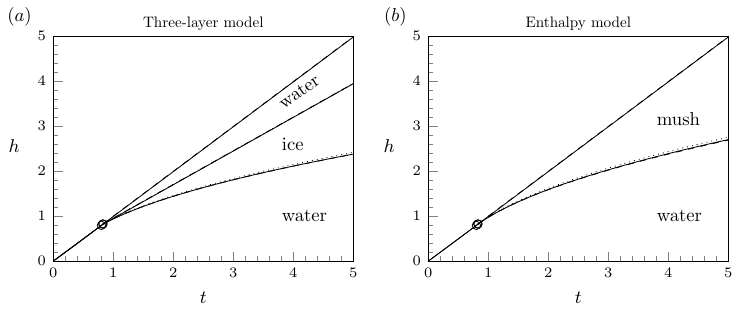}
    \caption{Typical evolution of the different heights for the two models using the baseline parameters as in \cref{tab:nd_param}. Solid lines are the full numerical prediction, dashed and dotted are the $O(\Pe^0)$ and $O(\Pe^1)$ solutions respectively for the $(a)$ three-layer model \eqref{eq:3Layer_sol}; and $(b)$ enthalpy model \eqref{eq:enthalpy_sol}.
    }
    \label{fig:evolve_three_enth}
\end{figure}
\begin{figure} 
    \centering
    \includegraphics{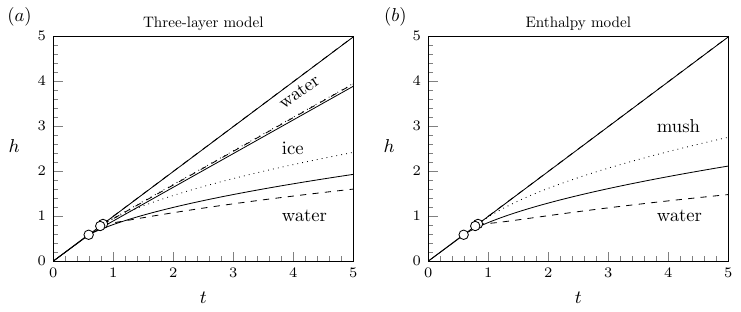}
    \caption{The same numerical experiment as in \cref{fig:evolve_three_enth} but now with the P\'{e}clet number 20 times the baseline value, $\Pe = 3.69$. We show $(a)$ the three-layer model; and $(b)$ the enthalpy model. The two-term asymptotic approximation, valid as $\Pe \to 0$, is shown dashed, while the leading-order is shown in dotted.  \label{fig:evolve_three_enth2}}
 \end{figure}   
We now show typical evolutions of the different heights for both the three-layer model and the enthalpy model in the cases of $\Pe = 0.185$ (\cref{fig:evolve_three_enth}) and $\Pe = 3.69$ (\cref{fig:evolve_three_enth2}). Firstly, in the insets $(a)$ of both figures, we plot the absolute heights of the three-layer model $\zwit, \ziwt,\ztop$ as defined earlier in \eqref{eq:height_absolute}. Similarly, in insets $(b)$ of both figures, we show the total height of the enthalpy method $\htotal$ and the extracted internal water layer $\hw$. In all cases, the asymptotic solutions to $O(\Pe^0)$ and $O(\Pe^1)$ are presented. For the three-layer model, asymptotic approximations of layer heights are given by \eqref{eq:h0_sol}--\eqref{eq:hs1_sol}; for the enthalpy model, the approximations are given in \eqref{eq:he0_sol}--\eqref{eq:hm1_sol}.

For the parameters we have investigated, the evolution of the system is typically qualitatively similar: in each figure, the single water layer grows until some critical height, $h^*$, and time, $t^*$, followed by the growth of the ice/water layers. The agreement with both the $O(\Pe^0)$ and the two-term ($O(\Pe^1)$, $h^{(0)}+\Pe h^{(1)}$) asymptotic approximation is excellent for the baseline value of $\Pe=0.185$. For example, at $t = 5$ there is $<1$\% error between the numerical and $O(\Pe^1)$ solutions for both the enthalpy and three layer models, where the asymptotic curve (dashed) is nearly visually indistinguishable from the full numerics (solid). However, as we increase the P\'{e}clet number to twenty times from the baseline value of $\Pe=3.69$, we observe that the correction term $h_1(t)$ overcompensates, resulting in a reduced water layer. We observe an increased reduction for the enthalpy compared to the three layer, as a result of dependence on $\Steff$ rather than $\St$. 

In \cref{fig:Pe_vs_t5}, we plot the numerical, leading-order solution, and two term solution at $t=5$, for different heights, against the P\'{e}clet number. The three-layer model water height is shown in \cref{fig:Pe_vs_t5}$(a)$ and the ice height in $(b)$. The enthalpy water height is plotted in \cref{fig:Pe_vs_t5}$(c)$ and the mush in $(d)$. As P\'{e}clet number increases, we observe divergence between the numerical and the asymptotic expansion, and thus show that neither the leading order nor two term approximation can accurately predict the evolution of the different layers for large P\'eclet. 

All existing ice crystal icing models assume quasi-steady-state heat transfer within the accretion layers, which is equivalent to the $O(\Pe^0)$ approximation. However, as shown in \cref{fig:Pe_vs_t5}, a deviation of more than 10\% occurs at $\Pe\approx0.5$, showing the inaccuracy of quasi-steady-state assumption. Note that $\Pe$ for the baseline is calculated using an impingement flux on the lower end from literature. The highest impingement flux given in \cref{tab:physicalconditions} leads to a P\'eclet number of 0.799, which will result in an even larger difference between the full transient and quasi-steady-state solutions.  Therefore, it is important to evaluate the P\'eclet condition when the quasi-steady-state assumption is taken. The effect of $\Pe$ and $\St$ on the critical time of freezing is explained in detail in \S{\ref{sec:num_tstar}}.

An apparent difference between the enthalpy model and the three-layer model is the thickness of the icing layer. As the enthalpy model is developed based on the physically-observed mixed-phase nature of the icing layer, it results in a much thicker icing layer compared to the pure ice layer in the three-layer model. The solutions of both models during the water-only stage are identical, as shown in \cref{fig:evolve_three_enth,fig:evolve_three_enth2} and \cref{tab:sweep_param}. However, after the onset of icing accretion ($t=t^*$), the water layer thickness of the enthalpy model develops faster, owing to the lower enthalpy level of the mushy layer and, hence, the modified Stefan condition on the interface with the mushy layer (see \eqref{eq:EnthStefan}).

The modified $\St$ is determined by the mushy phase enthalpy fraction, which is in turn determined by  the convective heat transfer parameter $\Bi$ and the impinging melt ratio $\MR$. If $\Bi$ increases for positive recovery temperatures, we would expect more heat transfer from convection, resulting in more water content. As we adjust melt ratio, smaller values reduce the time of freezing and significantly increase the ice layer. Table~\ref{tab:sweep_param} presents the effects of $\Bi$ and $\MR$ on $\hw^*$, $t^*$, $h(t=5)$, and $\hi(t=5)$ at the baseline $\Pe$. As $\MR$ is multiplied by 5, the onset of icing is delayed by 0.193 with a thicker $\hw^*$. As $\Bi$ increases, there will be eventually no icing due to enhanced heating on the top surface (see \cref{fig:Freeze_threshold}). In addition, the kinetic energy parameter $D$ will also affect the growth, as an increased $D$ will lead to greater kinetic energy in the heat flux, leading to more water content.

 \begin{figure}
     \centering
     \includegraphics{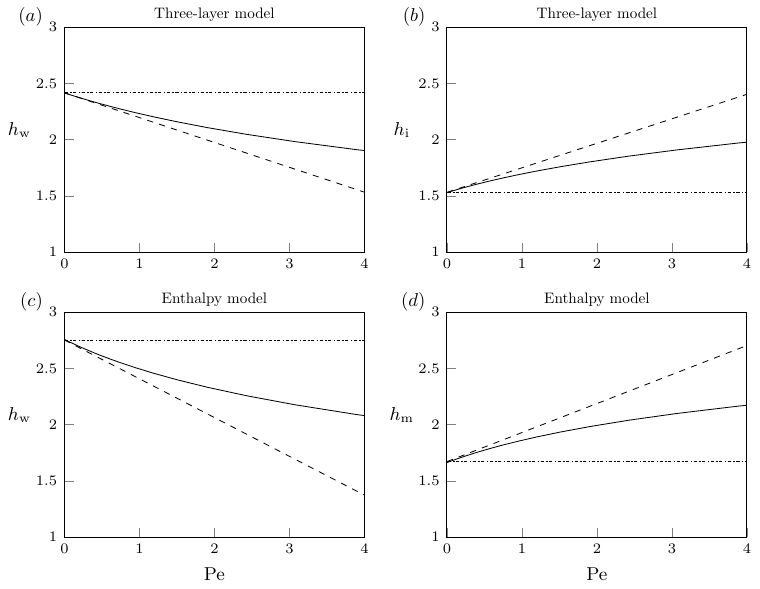}
     \caption{Plot of the numerical (solid lines), leading-order (dot-dash lines) and two term approximation (dashed lines) for various heights at $t=5$ vs the P\'{e}clet number, using the baseline parameters. The heights depicted are the $(a)$ Three-layer water; $(b)$ Three-layer ice; $(c)$ Enthalpy water; $(d)$ Enthalpy mush.}
     \label{fig:Pe_vs_t5}
 \end{figure}

\subsubsection{Bifurcation curves of the critical time of freezing} \label{sec:num_tstar}

In \cref{fig:Pevststar}$(a)$, we show the effect of the P\'{e}clet number on the critical time $t^*$ over a range of Stefan numbers, where the surface temperature reaches freezing. For small P\'{e}clet, $\Pe \ll 1$, the value of $t^*$ remains fairly constant. As the P\'{e}clet increases, there is a sharp decrease in the critical time. We also observe that agreement between the numerical and asymptotic solution for small P\'{e}clet is dependent on the Stefan number. The three Stefan numbers consist of our baseline value, and $\St_1 = 0.5, \; \St_2=5$, where all three are shown on \cref{fig:Freeze_threshold}. There is almost perfect agreement between the leading order and numerical solutions up until about $\Pe \approx 1$ for $\St =5$, $\Pe \approx 0.5$ for $\St = 1.61$, and $\Pe \approx 0.1$ for $\St = 0.5$. After these P\'{e}clet values, the critical time begins to decrease in the numerical solution leading to over prediction in the leading order asymptotic solution. We also note, that as the Stefan number is decreased, the range of agreement between leading order and numerical solution decreases. In \cref{fig:Pevststar}$(b)$, we plot the difference between the numerical solution $t^*$ and leading-order prediction ($t_0^*$ from \eqref{eq:critt0}) and compare this to the first order solution ($\Pe t_1^*$ from \eqref{eq:leadingfreezingeq}). We observe a linear trend for $\Pe < 1$ in the case of $\St =0.5$ and until $\Pe \approx 10$ for $\St =5$. We must note however, that both $t_0^*$ and $t_1^*$ are solved under the constant evaporation approximation from \S{\ref{sec:simplifymevconst}} and thus the discrepancy between the solid and dashed lines for small P\'{e}clet highlight the effect of this simplification. 

\begin{table}
\begin{tabular}{lp{1.2cm}p{1.2cm}p{1.2cm}p{1.2cm}p{1.2cm}p{1.2cm}p{1.2cm}p{1.2cm}}
\multicolumn{1}{c}{} & \multicolumn{4}{c}{\textsc{Three layer}}                 & \multicolumn{4}{c}{\textsc{Enthalpy}} \\
\multicolumn{1}{c}{} & \multicolumn{2}{c}{Bi (min, max)} & \multicolumn{2}{c}{$\MR$ (min, max)} & \multicolumn{2}{c}{Bi (min, max)} & \multicolumn{2}{c}{$\MR$ (min, max)} \\
 & 0.003 & 1.138 & 0.05 & 0.25 & 0.003 & 1.138 & 0.05 & 0.25 \\ \hline
$\hw^*$  & 0.776   & NA & 0.683  &  0.874  & 0.776   & NA & 0.683  &  0.874 \\
$t^*$   & 0.784 & NA  &  0.690 & 0.883 & 0.784 & NA  &  0.690 & 0.883  \\
$\hw(t=5)$  & 2.373  & 4.970  & 2.367  & 2.382 & 2.636  & 4.970 & 2.492 &2.782\\
$\hi(t=5)$ &  1.733 & NA & 2.188 & 1.368 & 1.851 & NA  & 2.235  &  1.534  
\end{tabular}
\caption{Variation of the Biot number - $\Bi$, and melt ratio - $\MR$, to typical minimum and maximum values, examining critical height, $h^*$ and time, $t^*$, for freezing to occur, as well as the height of the internal water and ice at a representative time, $t=5$. Note that in the case of the enthalpy model, we have extracted the ice component from the mush height as described in \eqref{eq:enth_ice}.  Other parameters are at the baseline value given in \cref{tab:nd_param}. Note that $\Bi=1.138 < \Bi_\text{crit}$ \eqref{eq:Bi_crit} for the baseline conditions and thus freezing would eventually occur.
}
\label{tab:sweep_param}
\end{table}

\begin{figure}
    \includegraphics{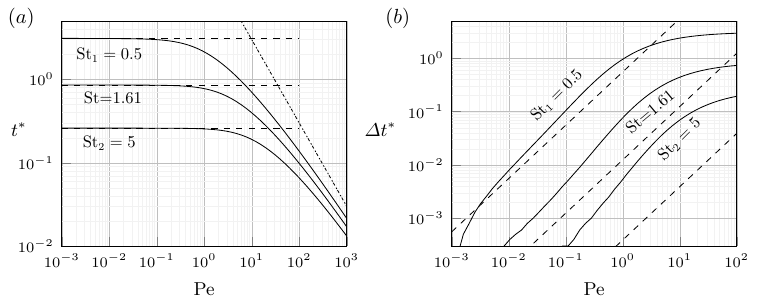}
    \caption{$(a)$ Loglog plot of $\Pe$ vs $t^*$. Solid lines give full numerical solutions while dashed lines is the leading-order asymptotic prediction from setting $T_0 = 0$ where we have $T_0$ from \eqref{eq:leadingfreezingeq}, and we solve for $h_0^*$ \eqref{eq:crith0} and $t_0^*$ \eqref{eq:critt0} The scaling is also confirmed to be $t^* = O(1/\Pe)$ for large $\Pe$ which is shown by the dash-dotted line. $(b)$ Loglog plot of $\Pe$ vs $t^*-t_0^{*}$. Solid lines give full numerical solutions, while dashed lines are the first order correction $t_1^{*}$. We calculate $t_1^{*}$ from \eqref{eq:firstordertimefre}. Both use the baseline parameters found in \cref{tab:nd_param} but adjust the P\'{e}clet and Stefan numbers. We sweep the P\'{e}clet number over three values of the Stefan number: our baseline value of $\St=1.68$ along with $\St_1=0.5$ and $\St_2=5$, which were used previously in \cref{fig:Freeze_threshold}.}
    \label{fig:Pevststar}
\end{figure} 

\section{Conclusions}\label{sec:conclusions}
Motivated by a previous three-layer formulation for modelling ice-crystal icing, in this work, we have developed a novel alternative that uses enthalpy to allow the simulation of mixed-phase icing. Our model shares similarities with a formulation presented in the computational experiments of \cite{malik2023experimental, Malik2024}; our analysis  carefully performs the analysis for the one-dimensional growth of the early mix-phase ice accretion layer for a warm substrate. The enthalpy model is compared to an existing three-layer model \citep{bucknell2019three}, which is also reformulated, non-dimensionalised, and studied more rigorously. In its simplest form, the icing problem is controlled by a group of non-dimensional parameters, including Pe, St, Mr, Bi, $\m_\text{ev}$, L, D which categorise regions of freezing and determine accretion growth.

Asymptotic solutions in the limit of $\Pe \to 0$ are derived, and these compare favourably to the numerical solutions showing the evolution of the accretion in both models. For $\Pe\lessapprox 0.5$, the asymptotic approximations present good agreement with numerical solutions. As the P\'eclet number increases, we observe divergence between the numerical and the asymptotic expansion; in this regime, neither the leading-order nor two-term approximation can accurately predict the evolution of the different layers. Prior works of the general ice-crystal icing problem have not performed asymptotic analysis to this level of detail; these analytical solutions also show clearly the effects of modifying parameters such as Bi and St. The comparison between the asymptotic and numerical solutions shows that the assumption of quasi-steady-state heat transfer within accretion layers, which was widely used in existing icing models, can only be valid at $\Pe\lessapprox 0.5$. 

\par
As compared to the three-layer model, the enthalpy model presents a thicker mushy accretion layer relative to the pure ice accretion of the three layer. The thickness of this layer, within the enthalpy model, is less than the sum of the ice and surface-water layers within the three-layer model. The enthalpy model predicts a greater internal water thickness, driven by the modified Stefan number determined by the enthalpy fraction. In the three-layer model, this difference is attributed to the application of an \emph{ad hoc} melting condition imposed on the top water layer by \citealt{bucknell2019three}, and a resulting thicker surface-water layer. Such differences are enhanced at large P\'{e}clet numbers.

\section{Discussion}
\label{sec:discussion}

In performing this work, the authors developed a significant appreciation for the hidden complexity of the three-layer model of ice crystal icing introduced in the recent work of \cite{bucknell2019three}. Although this layered model is only one of many reductions considered in the recent literature (see \emph{e.g} review by \cite{yamazakiReviewCurrentStatus2021}), it has found some success in providing a sufficiently simple conceptual model of the icing process that can form the basis of more complicated geometries and setups. To some extent, the complexity of the three-layer model is unavoidable: it must account for the multiple surface heat fluxes, which includes the effects of convection, evaporation, melting/freezing, sensible heat, and kinetic energy. Within a real aeroengine geometry, which is the primary motivator for this research, such contributions would be provided via flow analysis of the engine; therefore flexibility in specifying such fluxes is paramount. One of the contributions of this work has been to present the three-layer model in a more systematic and clear way, where the underlying assumptions are more transparent. The importance of non-dimensionalisation in highlighting key effects has also been stressed, and our survey of typical parameter sizes is useful for modelling. We have been able to develop asymptotic analysis in the limit of $\Pe \to 0$, which provides simple expressions for water and ice growth rates.

Previously, a quasi-steady approach was taken to model the water layer \citep{bucknell2018ice,connolly2021ice} (equivalent to considering $\Pe = 0$). In our work, we have more carefully justified under what conditions this assumption is valid, and we explored how the system is affected by the adjustment of further nondimensional parameters. Key behaviours, such as the time of freezing and the corresponding water thickness, are characterised numerically and asymptotically. 

The complexity of the three-layer model, along with its inability to model situations of mixed ice-water phases, thus motivated the development of the enthalpy model in this work. The enthalpy model is able to deal with conditions where the substrate temperature is above freezing, and can account for a ``mushy” layer as has been observed in icing under warm conditions. The framework is also computationally advantageous in that the temperature field can be solved within the entire domain, including both ice and water, and without requiring evolution of the interior interfaces. Our work also highlights the importance of the ``balancing enthalpy", i.e. the enthalpic value for which the heat flux is zero (cf. \eqref{eq:Estar}). In \cref{sec:Enth_interpret}, we demonstrated the connections between the balancing enthalpy and the previous melt/freeze rates established in \cite{bartkus2018evaluation,bucknell2019three}.

Following prior work by \emph{e.g.} \cite{bucknell2019three}, we have focused on the formulation and analysis of simplified one-dimensional dynamics for the three-layer and enthalpy models. In terms of replicating real engine environments, a number of extensions should be examined. For example, as a means of comparison to prior models, we have used assumptions of perfect thermal contact and infinite thermal capacity, \emph{e.g.} imposing $T(0, t) = \Tsubs$. However, there has been experimental evidence that suggests that it is important to consider the response of the substrate, itself, which may cool down due to particle impingement \citep{bartkus2018evaluation,currie2020physics, connolly2021ice,malik2023experimental}. Other important effects neglected include considerations of shear and pressure on water runback and erosion. There is a need for further analysis in realistic engine conditions that will establish thresholds for early-time icing, as functions of shear stress, mass flux, substrate temperature, and so forth.

Currently, the present authors are preparing an extension of this work that studies ice-crystal icing with two-dimensional dynamics, i.e. where the water and ice layers are allowed to be functions of a substrate coordinate. In the two-dimensional case, two substantial complexities are introduced. First, the temperature or enthalpy fields now require the solution of the two-dimensional diffusion equation within the ice/water regions. Second, the advection and thin-film dynamics must be considered, instead of \eqref{eq:nonDimEnth} and \eqref{eq:nonDimsurfaceevolve}; such dynamics capture the effects of water runback, shear, and pressure  \citep{Myers2002b}. Such thin-film models have also been considered by \emph{e.g.}  \cite{wright2015recent,currie2020physics,connolly2021ice}; like the detailed analysis between layered and enthalpy models presented here for the one-dimensional case, two-dimensional dynamics introduce significant and fascinating complexity. 

Beyond consideration of the water/ice phases itself, further analyses of the effects of particle impact and erosion are crucial. Previously, such effects have been incorporated in computational studies via phenomenological or empirical laws. While there have been advances in our understanding of ice-particle impact \citep{senoner2022ice, reitter2022impact}, detailed analyses of \emph{e.g.} the conditions for which an ice crystal adheres to a thin water film or solid surface in a high-speed flow remains challenging. We highlight work done by \emph{e.g.} \cite{hicks2011skimming, jolley2021particle, jolley2023interactions} on understanding ice-particle impact on surfaces; the work of \emph{e.g.} \cite{cimpeanu2018early,fudge2023drop} may be helpful in formulating mixed-phase problems where there are both water, ice, and air phases.

\mbox{}\par
{\bf \noindent Funding}. TP acknowledges support from the EPSRC Centre for Doctoral Training in Statistical Applied Mathematics at Bath (EPSRC grant no. EP/S022945/1). JS and PHT acknowledge support from EP/V012479/1, and JS is additionally supported by EP/W522491/1. \\

{\bf \noindent Declaration of interests}. The authors report no conflict of interest.

\appendix
\section{Further details for the numerical computation of the enthalpy model \label{sec:flag_Enth}}
In order to solve our system given by \eqref{eq:num_system}, we need to compose the matrix $\mathbf{M}$ given in \eqref{eq:matrix_M}.
Firstly, we construct the matrices $\mathbf{T}$ and $\mathbf{S}$ which fill up the upper half of $\mathbf{M}$, by substituting the finite difference stencils into our enthalpy equation \eqref{eq:Enth_PDE_coord},
\begin{equation}
\begin{aligned}
        & \bigg( \Pe\bigg[1 + \frac{\xi}{\hw}\frac{\mathrm{d}\hw}{\mathrm{d}t} \bigg]+ \frac{2\Delta t}{\hw^2 \Delta \xi^2} \bigg) T_{j}^{n+1} \\
        &-  \bigg(\frac{\Delta t}{\hw^2 \Delta \xi^2}\bigg) T_{j-1}^{n+1} -  \bigg( \Pe\bigg[\frac{\xi}{\hw}\frac{\mathrm{d}\hw}{\mathrm{d}t} \bigg]+ \frac{\Delta t}{\hw^2 \Delta \xi^2} \bigg)T_{j+1}^{n+1} \\
        &+ \bigg( \Pe\bigg[1 + \frac{\xi}{\hw}\frac{\mathrm{d}\hw}{\mathrm{d}t} \bigg]\bigg) S_{j}^{n+1} - \Pe\bigg(\frac{\xi}{\hw}\frac{\mathrm{d}\hw}{\mathrm{d}t} \bigg) S_{j+1}^{n+1}  = \Pe \; E_{j}^{n}.
\end{aligned}
\end{equation}
 As discussed in \eqref{eq:sol_vector}, $T_j^n$ and $S_j^n$ are the temperature and enthalpy at the $n^\text{th}$ time step at the location $\xi = \xi_j$. The lower half of the matrix $\mathbf{M}$ \eqref{eq:matrix_M} can be decomposed into the smaller matrices $\mathbf{A}$ and $\mathbf{B}$, which consist of ones on the main diagonal and are dependent on the flag regime. In addition, $\mathbf{A} + \mathbf{\hi} = \mathbf{I}$, where $\mathbf{I}$ is the identity matrix.
  Finally, we have the flag vector given by 
\begin{equation}
    \mathbf{F} = [F_{1}, \; F_{2}, \; \hdots\; F_{N}],
\end{equation}
for which the corresponding entries, $F_i$, indicate the substance at location $\xi = \xi_i$. 

\subsection{Flag updates}
 Each flag entry $F_i$ takes a value if $0$, $1$, or $2$, corresponding to whether the corresponding substance is ice, water, or mush, respectively. We consider the different cases and how it affects $\mathbf{A}, \mathbf{\hi}$ and the values of $\phi$ in $\mathbf{f}$.

\begin{enumerate}[label=(\roman*),leftmargin=*, align = left, labelsep=\parindent, topsep=3pt, itemsep=2pt,itemindent=0pt ]
     \item $F_i = 0 \Rightarrow$ \emph{Ice}  
\newline
There is no latent heat contribution for the enthalpy. This results in $\phi_i = 0, \; A_{i,i} = 0, \; B_{i,i} = 1$. Intuitively, this is setting $S_i = 0$.
     \item $F_i = 1 \Rightarrow$ \emph{Water} 
\newline
There is a latent heat contribution for the enthalpy. This results in $\phi_i = \St/\Pe, \; A_{i,i} = 0, \; B_{i,i} = 1$. Intuitively, this is setting $S_i = \St/\Pe$.
     \item $F_i = 2 \Rightarrow$ \emph{Mushy} 
     \newline
 The temperature is zero, resulting in $S \in [0, \St/\Pe]$. This results in $\phi_i = 0, \; A_{i,i} = 1, \; B_{i,i} = 0$. Intuitively, this is setting $\Ti = 0$.
 \end{enumerate}

 After each time step, we check the flag, examining the consistency of the solution. If the solution is inconsistent, we change the flag updating the matrices and run again. This is done in the following way, by comparing the flag with the temperature/latent heat at each grid point:
\begin{enumerate}[label=(\roman*),leftmargin=*, align = left, labelsep=\parindent, topsep=3pt, itemsep=2pt,itemindent=0pt ]
     \item ``Warm ice" $(F_i = 0 \land \Ti > 0)$ \newline
     The flag indicates the ice regime but the temperature is positive which is inconsistent. We set the flag to the mushy regime ($\Rightarrow F_i = 2)$.
     \item ``Freezing water" $(F_i = 1 \land \Ti < 0)$ \newline
     The flag indicates the water regime but the temperature is negative which is inconsistent. We set the flag to the mushy regime ($\Rightarrow F_i = 2)$.
     \item ``Negative latent heat" $(F_i = 2 \land S_i < 0)$ \newline
     The flag indicates the mushy regime but the latent heat contribution is negative which is inconsistent. We set the flag to the ice regime ($\Rightarrow F_i = 0)$.
     \item ``Super latent heat" $(F_i = 2 \land S_i > \St/\Pe)$ \newline
     The flag indicates the mushy regime but the latent heat contribution exceeds the threshold which is inconsistent. We set the flag to the water regime ($\Rightarrow F_i = 1)$.
 \end{enumerate}
This process is iterated at each time step until consistency is reached. Only $\mathbf{A}, \mathbf{\hi}$ and the values of $\phi$ in $\mathbf{f}$ need to be changed in this process.

\section{Evaporative mass flux} \label{sec:evap}

\noindent In \eqref{eq:EMMcontinuity} and \eqref{eq:Dimsurfaceevolve}, we assume that the evaporative mass flux, $\dot{m}_\text{ev}$, 
evaporation is determined by the following formula: 
\begin{equation}\label{eq:Appmevformula}
\m_{\text{ev}}(T) = \left[\frac{h_{\text{tc}}}{P_0 c_\text{a}Le^{1-b}}\frac{M_\text{w}}{M_\text{a}}\right](P_{\text{vap,sat,surf}}(T)-P_{\text{vap,sat},\infty}), 
\end{equation}
where $h_\text{tc}$ is the heat transfer coefficient, $P_0$ is the air pressure, $c_\text{a}$ is the heat capacity of air, $Le$ is the Lewis number, $b$ is a coefficient based on the Chilton-Colburn heat-mass transfer analogy, $M_\text{w}, M_\text{a}$ are the molar mass of water and air respectively, $P_{\text{vap,sat,surf}}$ is the vapor saturation pressure at the surface, $P_{\text{vap,sat}}(T_\text{surf})$, and $P_{\text{vap,sat},\infty}$ is the vapour saturation pressure at $T_{\infty}$ multiplied by the relative humidity, $P_{\text{vap},\infty} = RH_\infty P_{\text{vap,sat}}(T_\infty)$.  The above is taken from eqn (A8) in \cite{bucknell2019three}. 

Various equations can be used to calculate the vapour saturation pressure, $P_\text{vap,sat}$. Here Hyland \& Wexler's equation is used \citep{wexler1983thermodynamic}, and can be written as
\begin{equation}
    P_\text{vap,sat}(T)= \begin{cases}
        \displaystyle \exp \left[c_4\log (T + T_K)+ \sum_{i=-1}^3 c_i (T + T_K)^i \right]  & T \geq 0, \\
        \displaystyle \exp \left[d_4\log (T + T_K)+ \sum_{i=-1}^3 d_i (T + T_K)^i \right] & T < 0,
    \end{cases}
    \label{eq:Hyland}
\end{equation}
where $T_K=273.15$ is the temperature shift from Celsius to Kelvin and the coefficients,  $c_i$ and $d_i$, for $i = -1, 0, 1, \ldots, 4$, are different for vapour over liquid water and solid ice, and are listed in \cref{tab:vapourcoeffs}.

\begin{table}
    \begin{tabular}{rS[table-format=6.0]>{\hspace{2pc}}S[table-format=3.1] >{\hspace{3pc}} S[table-format=3.3]>{\hspace{3pc}}S[table-format=3.3]>{\hspace{4pc}}S[table-format=4.3]>{\hspace{2pc}}S[table-format=3.]}
          & {$c_{-1}$} & {$c_0$} & {$c_1$} & {$c_2$} & {$c_3$}  & {$c_4$} \\ 
         Liquid water& -5800 & 1.391 & -4.864{$\times10^{-2}$} & 4.176{$\times10^{-5}$}   & -1.445{$\times10^{-8}$} & 6.545 \\[1em]
          & {$d_{-1}$} & {$d_0$} & {$d_1$} & {$d_2$} & {$d_3$}  & {$d_4$} \\
         Ice         & -5674 & 6.392  & -9.677{$\times10^{-3}$} & 6.221{$\times10^{-7}$} & 2.074{$\times10^{-9}$} & 4.163\\
    \end{tabular}
    \caption{Coefficients from \cite{wexler1983thermodynamic} used to calculate vapour saturation pressure in \eqref{eq:Hyland}. }
    \label{tab:vapourcoeffs}
\end{table}

Previously, \cite{myers2001extension} provided a linear fit for the evaporative flux, while \cite{bucknell2019three} used a piecewise linear fit for the saturation pressure. By writing a linear fit for the saturation pressure, this leads to a linear fit for the nondimensional evaporative flux $\m_\text{ev}$ for $T \in [0,10] ^\circ C$, which is then used in \S{\ref{sec:asympt}}. This is 
\begin{equation}
\begin{aligned} \label{eq:nondimev}
    \m_\text{ev}(T) &\approx \left[\frac{h_{\text{tc}}}{P_0 c_\text{a}Le^{1-b}}\frac{M_\text{w}}{M_\text{a}}\right](\tilde{e}_1 + \tilde{e}_2 T - RH(\tilde{e}_1 + \tilde{e}_2 T_\infty)), \\
    \m_\text{ev}'(T) &\approx \left[\frac{h_{\text{tc}}}{P_0 c_\text{a}Le^{1-b} \mimp \Trec}\frac{M_\text{w}}{M_\text{a}}\right](\tilde{e}_1 - RH(\tilde{e}_1 + \tilde{e}_2 T_\infty) + \tilde{e}_2 T), \\
    \m_\text{ev}'(T) &\approx \evI + \evII T,
\end{aligned}
\end{equation}
where $\evI = 0.003$ and $\evII=0.0017$ and are both nondimensional. This is accurate to within 7\% over the specific range. Since for the most part use of \eqref{eq:nondimev} does not structurally yield so much more simplification than the full model \eqref{eq:Appmevformula}, we have used the full model in our numerical work. 

\providecommand{\noopsort}[1]{}\providecommand{\singleletter}[1]{#1}%

\end{document}